\documentclass[manuscript]{aastex}

\usepackage{natbib}

\slugcomment{To appear in The Astrophysical Journal}
\shorttitle{{\it Archeoseismology}: bridging stellar and Galactic astrophysics}
\shortauthors{Casagrande et al.}

\newcommand{\teff}{T_{\rm{eff}}}
\newcommand{\logg}{\log g}
\newcommand{\feh}{\rm{[Fe/H]}}
\newcommand{\fbol}{\mathcal{F}_{\rm{Bol}}}
\newcommand{\ebv}{E(B-V)}

\begin{document}

\title{Str\"omgren survey for Asteroseismology and Galactic Archaeology: let 
the SAGA begin\footnote{Based on 
observations made with the Isaac Newton Telescope operated on the island of 
La Palma by the Isaac Newton Group in the Spanish Observatorio del Roque de 
los Muchachos of the Instituto de Astrof\'isica de Canarias.}}

\author{
        L.\,Casagrande     \altaffilmark{2,\dag},
        V.\,Silva Aguirre  \altaffilmark{3},
        D.\,Stello         \altaffilmark{4},
        D.\,Huber          \altaffilmark{5,6},
        A.M.\,Serenelli    \altaffilmark{7},
        S.\,Cassisi        \altaffilmark{8},
        A.\,Dotter         \altaffilmark{2},
        A.P.\,Milone       \altaffilmark{2},
        S.\,Hodgkin        \altaffilmark{9},
        A.F.\,Marino       \altaffilmark{2},
        M.N.\,Lund         \altaffilmark{3},
        A.\,Pietrinferni   \altaffilmark{8},
        M.\,Asplund        \altaffilmark{2},
        S.\,Feltzing       \altaffilmark{10},
        C.\,Flynn          \altaffilmark{11},
        F.\,Grundahl       \altaffilmark{3},
        P.E.\,Nissen       \altaffilmark{3},
        R.\,Sch\"onrich    \altaffilmark{12,13}
        K.J.\,Schlesinger   \altaffilmark{2}
        W.\,Wang           \altaffilmark{14}
      }

\altaffiltext{2}{Research School of Astronomy \& Astrophysics, Mount Stromlo 
                 Observatory, The Australian National University, ACT 2611, 
                 Australia}
\altaffiltext{3}{Stellar Astrophysics Centre, Department of Physics and 
                 Astronomy, Aarhus University, Ny Munkegade 120, DK-8000 
                 Aarhus C, Denmark}
\altaffiltext{4}{Sydney Institute for Astronomy (SIfA), School of Physics, 
                 University of Sydney, NSW 2006, Australia}
\altaffiltext{5}{NASA Ames Research Center, Moffett Field, CA 94035, USA} 
\altaffiltext{6}{SETI Institute, 189 Bernardo Av., Mountain View, CA 94043, USA}
\altaffiltext{7}{Institute of Space Sciences (IEEC-CSIC), Campus UAB, 
                  Fac.~Ci\'encies, Torre C5 parell 2, E-08193 Bellaterra, Spain}
\altaffiltext{8}{INAF-Osservatorio Astronomico di Collurania, via Maggini, 
                  64100 Teramo, Italy}
\altaffiltext{9}{Institute of Astronomy, Madingley Road, Cambridge CB3 0HA, UK}
\altaffiltext{10}{Lund Observatory, Department of Astronomy \& Theoretical 
                  Physics, Box 43, SE-22100, Lund, Sweden}
\altaffiltext{11}{Centre for Astrophysics and Supercomputing, Swinburne 
                 University of Technology, VIC 3122 Australia}
\altaffiltext{12}{Department of Astronomy, The Ohio State University, 140 West 
                 18th Ave., Columbus, OH 43210-1173, USA; Hubble Fellow}
\altaffiltext{13}{Rudolf-Peierls Centre for Theoretical Physics, University of 
                  Oxford, 1 Keble Road, OX1 3NP, Oxford, United Kingdom}
\altaffiltext{14}{National Astronomical Observatories, Chinese Academy of 
                  Sciences, Beijing 100012, China}
\altaffiltext{\dag}{Stromlo Fellow}
\email{luca.casagrande@anu.edu.au}

\begin{abstract}

Asteroseismology has the capability of precisely determining stellar properties 
which would otherwise be inaccessible, such as radii, masses and thus ages of 
stars. When coupling this information with classical 
determinations of stellar parameters, such as metallicities, effective 
temperatures and angular diameters, powerful new diagnostics for Galactic 
studies can be obtained. The ongoing Str\"omgren survey for Asteroseismology 
and Galactic Archaeology (SAGA) has the goal of transforming the 
{\it Kepler} field into a new benchmark for Galactic studies, similarly to the 
solar neighborhood. 
Here we present first results from a stripe centred at Galactic longitude 
$74^{\circ}$ and covering latitude from about $8^{\circ}$ to $20^{\circ}$, which 
includes almost 1000 K-giants with seismic information and the benchmark open 
cluster NGC\,6819. We describe the coupling of classical and seismic 
parameters, 
the accuracy as well as the caveats of the derived effective 
temperatures, metallicities, distances, surface gravities, masses, and radii. 
Confidence in the achieved precision is corroborated by the detection 
of the first and secondary clump in a population of field stars with a 
ratio of 2 to 1, and by the negligible scatter in the seismic distances 
among NGC\,6819 member stars. An assessment of the reliability of stellar 
parameters in the Kepler Input Catalogue is also performed, and the impact of 
our results for population studies in the Milky Way is discussed, along with 
the importance of an all-sky Str\"omgren survey.

\end{abstract}


\section{Introduction}\label{sec:intro}

The study of a sizeable number of individual stars in the Milky Way enables us 
to directly access different phases of its formation and evolution, in a 
fashion which is still hardly achievable for external galaxies. For obvious 
reasons, 
stars in the vicinity of the Sun have been preferred targets for this 
purpose, starting from the very first \citep[e.g.,][]
{gliese57,wallerstein62,vandenBergh62,els62} until the most recent photometric 
and spectroscopic investigations of the Milky Way, for which {\it Hipparcos} 
astrometric distances are also available out to 
$\simeq 100$~pc \citep[e.g.,][]{nordstrom04,reddy06,fb08,c11,ad13,b13}.
Properties of stars in the solar neighborhood, in particular ages and 
metallicities, are still the main constraint for Galactic chemo(dynamical) 
models 
\citep[e.g.,][]{cescutti07,ralph09a,kn11,m12,wz13} and provide important 
clues to understanding the main processes at play in galaxy formation and 
evolution \citep[e.g.,][]{pilk12,bird13}.

This sort of study is now fostered by a number of spectroscopic and 
photometric surveys, targeting different and fainter components of the Milky 
Way (e.g.,~``RAVE'' \citealt{steinmetz06}; ``SDSS-SEGUE'' \citealt{ya09}; 
``SDSS-APOGEE'' \citealt{ahn13}; ``VVV'' \citealt{minniti10}; ``Gaia-ESO'' 
\citealt{gge}; ``GALAH'' \citealt{ken}), although astrometric distances for 
the targets in these surveys will not be available until the {\it Gaia} 
spacecraft releases its data \citep{lp96:gaia,p01:gaia}. 
A common feature of all past and current stellar surveys is that while it is 
relatively straightforward to derive some sort of information on the chemical 
composition of the targets observed (and in many cases even detailed 
abundances), that is not the case when it comes to masses, radii, 
distances and, in particular, ages. Even when accurate astrometric distances 
are available to allow comparison of stars with isochrones, the derived ages 
are still highly uncertain, and statistical techniques are required to avoid 
biases \citep[e.g.,][]{pont04,Jorgensen05}. Furthermore, isochrone 
dating is meaningful only for stars in the turnoff and subgiant phase, where 
stars of different ages are clearly separated on the HR diagram. This is 
in contrast, for example, to stars on the red giant branch, where isochrones 
with 
vastly different ages can fit equally well observational constraints such as 
effective temperatures, metallicities and surface gravities within their errors
\citep[e.g.,][for a review]{soderblom}. Thus, alternative ways to precisely 
determine masses and radii of stars are the only way forward.

By measuring oscillation frequencies in stars, asteroseismology allows us to 
determine fundamental physical quantities, in particular masses and radii, 
which otherwise would be inaccessible in single field stars, and which can be 
used to obtain information on stellar distances and ages \citep[e.g.,][for a 
review]{cm13}. 
Individual frequencies can yield an accuracy of just a few percent on 
those parameters but their exploitation is more demanding, both 
observationally and theoretically \citep{vsa13}. Global asteroseismic 
observables (see Section \ref{sec:astero}) on the contrary are easier to 
detect and analyze, yet able to provide the aforementioned parameters for a 
large number of stars with an accuracy that is still much better than 
achievable by isochrone fitting in the traditional sense.
Thanks to space-borne missions such as {\it CoRoT} \citep{corot06} and 
{\it Kepler} \citep{gilli10} average oscillation frequencies are now robustly 
detected in more than 500 main-sequence and subgiant stars, and over $13,000$ 
giants \citep[e.g.,][]{derid09,verner11,stello13}.
Asteroseismology is thus emerging as a new tool for studying stellar 
populations, and initial investigations in this direction have already been 
done \citep[][]{chap11,miglio13}. 
However, until now asteroseismic studies of stellar populations had only coarse 
information on ``classical '' stellar parameters such as $\teff$ and $\feh$. 
Coupling classical parameters with seismic information not only improves the 
seismic masses and ages obtained for stars \citep[][]{lm09,chap13}, but also 
allows us to address important questions in stellar and Galactic evolution.

To fully harvest the potential that asteroseismology brings to these studies, 
classical stellar parameters are vital, yet unavailable for a 
large sample of stars with detected oscillations. The main purpose of the 
Kepler Input Catalog \citep[KIC,][]{kic} was to separate dwarfs from giants, 
and it is therefore inadequate for high precision stellar and Galactic studies 
due to significant biases in its stellar parameters 
\citep[e.g.,][]{mz11,pi12,thy12}. APOGEE 
will eventually provide parameters for thousands of {\it Kepler} giants, and 
similarly the Gaia-ESO Survey and GALAH for the {\it CoRoT} fields. The 
advantages of the spectroscopic surveys are obvious, yet 
photometry still offers a powerful and complementary alternative.

Here, we present the first results from the ongoing Str\"omgren survey for 
Asteroseismology and Galactic Archaeology (SAGA) in the {\it Kepler} 
field. The goals of our 
photometric survey are manifold. First, the Str\"omgren system was envisaged 
with the ultimate purpose of studying Galactic stellar populations, and 
designed to provide reliable stellar parameters, in particular metallicities 
\citep[][]{stromgren63,stro87}. Thus, even compared to multi-fiber 
spectroscopy, wide-field Str\"omgren imaging is very efficient. It also has the 
advantage that no pre-selection is made on targets: all stars that fall in a 
given brightness regime across the instrument field of view will be observed. 
This greatly simplifies the selection function, which in our case is 
essentially dictated by the {\it Kepler} satellite \citep[e.g.,][]{bata,fk13}.
Further, relatively faint magnitudes can be probed at high precision even on a 
small-class telescope, making it possible to readily derive metallicities over 
a large magnitude range. The {\it Kepler} seismic sample presented in this 
work can thus be used as reference e.g.~for assessing the accuracy at which 
stellar parameters can be obtained for stars having only photometric 
measurements, such as the planet host stars we observe in SAGA. At the same 
time, the sensitivity of Str\"omgren colors to metallicity, coupled with 
seismic ages, help immensely to calibrate other metallicity and age dating 
techniques, thus creating powerful new tools to study more remote Galactic 
populations than previously possible.
Bearing in mind the expected precision at which {\it Gaia} will deliver 
astrophysical parameters for stars of different brightness \citep{liu12},
the above science case highlight the importance of having an all-sky 
Str\"omgren survey going fainter than any of the current large spectroscopic 
surveys \citep{wang}.

The SAGA survey also represents a natural extension of the Geneva Copenhagen 
Survey 
\citep[GCS,][]{nordstrom04}, the only all-sky Str\"omgren survey currently 
available, and a benchmark for Galactic studies. Similar to our latest 
revision of the GCS \citep{c11}, we combine Str\"omgren metallicities with 
broad-band photometry to obtain effective temperatures ($\teff$) and 
metallicities ($\feh$) for all our targets via the Infrared Flux Method (IRFM).
This facilitates the task of placing SAGA and GCS on the same scale. 
However, there are marked differences between the two surveys: the GCS is 
an all-sky, shallow survey limited to 
main-sequence and subgiant stars closer than $\simeq 100$~pc ($40$~pc volume 
limited). The {\it Kepler} targets observed by SAGA are primarily giants 
located between $\simeq1$ and $\simeq6$~kpc in a specific region of the 
Galactic disk (Figures \ref{f:fov} and \ref{f:trigo}).
The use of giants as probes of Galactic Archaeology is possible since it is 
relatively straightforward to derive ages for these stars once classical 
stellar parameters are coupled with asteroseismology. This was not the case 
for the GCS, where isochrone fitting was used, and thus limited to 
main-sequence and subgiant stars with known astrometric distances. On the other 
hand, stars in the GCS have kinematic information, which is not available for 
the SAGA targets. The different distance ranges sampled by the GCS and SAGA 
makes them complementary: the stellar properties measured within the solar 
neighborhood in the former survey can be dynamically stretched across 
several kpc using kinematics. In contrast, the larger distance range 
sampled by the giants in SAGA provides {\it in situ} measurements of 
various stellar properties over $\simeq5$~kpc.

In this paper we present the observing strategy and data 
reduction of the Str\"omgren survey. After cross-matching this dataset with 
stars in the {\it Kepler} field with seismic information, we derive 
classical and seismic stellar parameters for almost 1,000 targets. 
Our approach is characterized by the powerful combination of 
classical and seismic parameters, to our knowledge the first of this kind, 
and a careful treatment of the errors, allowing us to derive 
self-consistent effective temperatures, distances, masses and radii with a 
typical precision of a few percent, in addition to metallicities as we 
discuss in more detail later. With these parameters in hand, stellar ages are a 
straightforward by-product, although more caution must be used: we leave their 
estimation to a subsequent publication. 

\section{The Str\"omgren survey}

The $uvby$ photometric system \citep{stromgren63} is designed for the 
determination of basic stellar atmospheric parameters through the color 
indices $(b-y)$, $m_1 = (v-b)-(b-y)$ and $c_1 = (u-v)-(v-b)$. The $y$ 
magnitude is defined to be essentially the same as Johnson $V$ for stars 
other than M-type, and the color $(b-y)$ is relatively unaffected by line 
blanketing and thus well suited for measuring effective temperatures (although 
in this study we will determine $\teff$ in a more fundamental way, by 
using broad-band photometry in the IRFM, see Section \ref{sec:teff}).
The $m_1$ index is designed to measure the depression 
owing to metal lines around $4100$~\AA~ ($v$ band), and hence is suitable to 
infer the metallicity of a variety of stars. Finally, the $c_1$ index is 
designed to evaluate the Balmer discontinuity, which is a temperature indicator 
for B- and A-type stars and a surface gravity indicator for late-type stars, 
though for stars with lower or comparable temperature than the Sun it also 
carries metallicity information \citep[see e.g.,][and references therein]
{bessell05,onehag09,arna10}. 

\subsection{Observations and data reduction} \label{sec:sample}

Intermediate-band Str\"omgren $uvby$ photometry was obtained during seven 
nights with the Wide Field Camera (WFC) on the $2.5$-m Isaac Newton Telescope 
(INT) on La Palma. The camera comprises four $2\rm{k} \times 4\rm{k}$ thinned 
EEV CCDs; the coverage of each of the chips is $22.8 \times 11.4$ arcmin for 
a total field of view of $34 \times 34$ arcmin. Pixel size is $13.5$ microns, 
corresponding to $0.33$ arcsec/pixel, thus making the instrument ideal for 
wide field optical imaging surveys \citep[e.g.,][]{drew05,groot09,kis}.

Of the seven nights awarded during period 2012A, only four provided useful 
data due to bad weather on the other nights. Typical seeing during the four 
good nights was about $1.1$ arcsec, implying a typical FWHM of the 
stellar PSF of $3 \textrm{--} 4$ pixels.

Str\"omgren standard stars were chosen from the list of \cite{sn88}, which is 
carefully tied to the system used by \cite{olsen83} and underlying the GCS 
used for our previous investigation of stellar properties in the Galactic 
disk \citep{c11}. We use secondary rather than primary Str\"omgren standards as 
the latter are too bright to 
observe with a $2.5$-m telescope. Throughout each of the four photometric 
nights 10 standard stars were usually observed multiple times over a large 
span of magnitude and color indices, and at different elevations in the sky, 
bracketing 
pointings on the {\it Kepler} field both in airmass and time. 

A stripe across the {\it Kepler} field was observed, covering Galactic latitude 
$7.6^{\circ} < b < 19.9^{\circ}$ and centred at Galactic longitude 
$l = 74^{\circ}$. 
About 60\% of the pointings were 
observed more than once on at least two different photometric nights. 
We chose a tiling strategy with an overlap of $\simeq3$ to $\simeq10$ 
arcsec between different pointings over the {\it Kepler} field, thus allowing 
to check that photometric zero-points are constant over the observed region. 
This strategy allows us to 
uniformly sample the Galactic disk, as encompassed by the field of view of 
{\it Kepler}. The lower limit on $b$ is set to avoid regions too close to 
the plane, where crowding and reddening make it more difficult to use 
photometric data. The choice of the Galactic longitude upon which our stripe 
is centered was dictated by the fact that while this region is only moderately 
affected by dust and obscuration, it also includes the open cluster NGC\,6819, 
which provides a useful benchmark, as discussed later. 

The purpose of the SAGA survey is to obtain accurate Str\"omgren photometry for 
stars in the magnitude range $10 \lesssim y \lesssim 14$, where 
{\it Kepler} is able to detect oscillations in most red giant stars 
\citep[e.g.,][]{Huber:2011be}.
For science pointings, a typical observing sequence comprised one short 
($\simeq 5-10$ sec) and one long ($30-60$~sec) exposure in $vby$, and two 
short ($\simeq30$~sec) and one long ($\simeq 120$~sec) in $u$, after which a 
$45$~arcsec dither was applied and the above cycle repeated. With this 
strategy we are able to achieve photometric errors $<0.03$~mag in all bands 
over the magnitude range of interest (see later Figure \ref{f:errors}), which 
is the accuracy needed to obtain robust parameters from Str\"omgren indices 
\citep[e.g.,][]{cala07}. However, useful stellar properties can still be 
derived down to $16^{\rm{th}}$ magnitude, a regime important for 
planet-candidate host stars \citep{Batalha:2013fg,bestkoi}.

The images of the {\it Kepler} field and the standard stars were pre-processed 
with 
the Wide Field Survey Pipeline provided by the Cambridge Astronomical Survey 
Unit \citep{int}. The operations applied to the images were de-biasing, 
trimming, flat-fielding and correction for non-linearity.

\subsection{Standard stars: photometric calibration}

Aperture photometry for standard stars was obtained with the task PHOT within 
the IRAF\footnote{IRAF is distributed by the National Optical Astronomy 
Observatory, which is operated by the Association of Universities for Research 
in Astronomy (AURA) under cooperative agreement with the National Science 
Foundation.} DAOPHOT package. For each star we explored different sizes of the 
aperture, ranging from 5 to 20 pixels, and found a value of 15 to be optimal. 
This roughly corresponds to $4\textrm{--}5 \times \rm{FWHM}$ of the typical 
stellar PSF, and it is ideal for well isolated standards as it is in our case 
\citep[e.g.,][]{ccdphot}. In order to transform instrumental (inst) into 
standard magnitudes we adopted the following equation for each of the
Str\"omgren filters ($i=u,v,b,y$)
\begin{equation}\label{eq:pho1}
m_{inst,i} = m_{i} + \epsilon_{1,i} X + \epsilon_{2,i} (m_{v}-m_{y}) + \epsilon_{3,i} T + \epsilon_{4,i},
\end{equation}
where $X$ and $T$ denote the airmass and time of that CCD exposure, and
$m_{v}-m_{y}$ is the color of the standard system 
\cite[e.g.,][]{harris81,gru02}. The 
terms $\epsilon_{1,i} \dots \epsilon_{4,i}$ were determined individually for 
each photometric night; 
cross checks for stars in common observed over different nights confirm 
these terms are stable to better than $0.01$~mag in each band. We found the 
$\epsilon_{3,i}$ time-term to be significant in $u$ and $v$, while essentially 
negligible in $b$ and 
$y$. For the sake of our discussion, Equation \ref{eq:pho1} can be rewritten 
highlighting only the color term 
\begin{equation}\label{eq:pho2}
m_{inst,i} = m_{i} + \epsilon_{2,i} (m_{v}-m_{y}) + k_{i},
\end{equation}
where $k_{i}$ now includes all previous airmass, time and zero-point terms.

As customary with CCD photometry, the transformation to the 
standard system is done using individual magnitudes rather than Str\"omgren 
indices, since each filter is observed separately and hence with a time delay. 
This is in contrast to four-channel photometry where observations for all four 
filters are obtained at the same time, also allowing observations during non 
photometric nights \citep[e.g.,][]{olsen83,sn88,melendez10}. Calibrating upon 
individual magnitudes has 
the advantage that we need not to worry that observations through the 
different filters are obtained at different airmasses for each standard star. 
Figure \ref{f:standards} shows the residual between our photometry and the 
standard stars: excellent agreement is obtained in all four filters with small 
scatter and no obvious trends as function of $(v-y)$. The same check using 
other color combinations also shows no trends.
 
\subsection{Science stars: photometric calibration}\label{sec:sspc}

Aperture photometry was done on each of the science images separately. 
A dedicated suite of scripts has been developed for this purpose, as described 
below. Coordinate lists for the images in each band were created using the 
IRAF task DAOFIND. Aperture photometry was done with apertures ranging from 5 
to 20 pixels in steps of 1 pixel. Flux and photometric errors (the {\tt merr} 
output in PHOT, essentially determined from photon statistics) of each source 
were extracted at all selected apertures. 
For each image we searched for the brightest stars above a predefined 
instrumental magnitude, having {\tt merr} $<0.015$~mag and separated from any 
other detected source by at least twice the maximum aperture\footnote{More 
precisely, by twice the annulus plus dannulus as defined in DAOPHOT.}. An 
iterative scheme was employed to move the brightness threshold to fainter 
magnitudes if less than 10 such stars were identified. These stars were then 
used to compute the curve-of-growth, which was found to remain essentially 
constant for apertures larger than 15 pixels. The aperture correction was then 
computed for each frame as the median of the flux ratio measured in an 
aperture of 15 and 5 pixels, using a $2.6\,\sigma $ clipping to remove 
outliers. The mean absolute deviation was used to define $\sigma$, which 
together with the median described above provides robust statistics on our 
definition of the aperture correction. In this way, we also determined for 
each magnitude the systematic error stemming from the uncertainty $\sigma$ in 
the aperture correction.
This procedure was run in an iterative mode, by checking in each frame the 
position of bright stars on the CCDs and the resulting curve-of-growth. 
In all instances, results were found to be robust, with no need for further 
human intervention, thus making it possible to run the entire procedure in 
batch mode. For each frame, astrometric solutions were computed using the 
``World Coordinate System'' (WCS) from the image header and correcting for 
field distortion. By cross-matching the position 
of several of the brightest targets against 2MASS, we conclude the astrometric 
precision for our coordinates is $0.2-0.3$~arcsec\footnote{See also http://www.ast.cam.ac.uk/$\sim$wfcsur/technical/astrometry/}.
In very few cases (less than 1\%) the astrometric solutions were found to 
differ from 2MASS 
coordinates with trends depending on the X and Y position on the CCD; those 
images were excluded from further analysis. 
For each object identified in each frame, the {\tt RA} and {\tt dec} distance 
with respect to all other objects identified in subsequent frames and filters 
was computed: the vast majority of distances cluster into an ellipse of axis 
$0.4$~arcsec in {\tt RA} and $0.3$~arcsec in {\tt dec}, which is thus the 
typical accuracy at which a given object is identified in different frames and 
filters. This search ellipse was used to match sources in different bands and 
frames throughout the remainder of the data reduction.

By matching sources in different filters, we are thus able to apply the 
photometric calibration to the standard system. Notice though, 
Equation \ref{eq:pho2} depends on the standard color $m_{v}-m_{y}$ 
which for science targets is not known beforehand. Therefore, we first solved 
for $m_{v}$ and $m_{y}$
\begin{equation}\label{eq:pho3}
\left( \begin{array}{c}
m_{v}\\
m_{y}
\end{array}\right) = \left( \begin{array}{cc}
1+\epsilon_{2,v} & -\epsilon_{2,v} \\
\epsilon_{2,y}   & 1-\epsilon_{2,y}
\end{array}\right)^{-1} \left( \begin{array}{c}
m_{inst,v} - k_v \\
m_{inst,y} - k_y 
\end{array}\right)
\end{equation}
such that Equation \ref{eq:pho2} could be applied to $m_{u}$ and $m_{b}$ 
afterwards. 
For each night, calibrated photometry for each frame in a given filter was 
cross-matched for stars in common with all other frames in the same filter. 
We then calculated the median offset with respect to all other images 
for each frame; zero-point differences were on average a few millimag (thus 
confirming the stability of our measurements on photometric nights), apart from 
occasional offsets spurring to several hundredths (or more) of a magnitude.
The median is essentially unaffected by the presence of strong outliers and an 
iterative scheme was employed to bring all frames to exactly the same 
zero-point, whichever of the following conditions was first satisfied: an 
average median offset with respect to all other frames in the same filter and 
in the same night reached below $0.005$~mag or if we reached at least 7 
iterations. This second condition was chosen to avoid the rare cases when the 
zero-point correction oscillates just above the $0.005$~mag threshold, since 
convergence within $0.01$~mag is always reached within 3 or 4 iterations at 
most.
This zero-point shift for each frame was included as a systematic 
uncertainty in all the photometric values of that given frame. 
Since Equations \ref{eq:pho2} and \ref{eq:pho3} depend on the standard 
$m_{v}-m_{y}$ color, those were reapplied after correcting for the zero-points 
determined above. 

The photometric error associated with each measurement was derived from the sum 
of squared residuals stemming from photon statistics, aperture correction 
uncertainty, and the zero-point shifts described above. If more than one 
measurement was available, then the weighted average (using the above 
photometric error as weight) was used and the standard deviation taken as 
measurement error. Based on these global photometric errors, we define 
fiducials 
for selecting stars with the most reliable photometry; those ridge-lines are 
shown in Figure \ref{f:errors}. A star is flagged to have good photometry 
only if the errors in all four bands fall below those ridge-lines.

Since we are interested in determining $uvby$ for all stars, we took $u$ band 
measurements (our faintest band), and searched for matches in all other 
filters. 
The coordinate of each object having Str\"omgren magnitudes was determined by 
computing the average coordinate from $uvby$ bands, using the standard 
deviation as the uncertainty. The final catalogue contains $29,521$ sources 
with $uvby$ measurements (all shown in Figure \ref{f:errors}). Changing the 
matching criterion to a circular search of radius 
1 arcsec returns essentially the same number of objects ($29,496$).

\subsection{Comparison with other Str\"omgren studies}

The only all-sky Str\"omgren survey is the GCS, which is largely based on the 
extensive work of \cite{olsen83,olsen84,olsen94a,olsen94b} upon which we ought 
to calibrate. Most of the stars in the GCS are brighter than our saturation 
limit and are sparsely distributed across the entire sky, resulting in no 
object in common with our observations. However, we can easily verify if our 
observations follow the mean loci of good Str\"omgren photometry in the Olsen 
system. In the left panel of Figure \ref{f:strom} we plot our measurements
in the $c_1$ vs $(b-y)$ plane, which discriminates between 
dwarfs and giants \citep[e.g.,][]{faria07,aden09}. Overplotted is the fiducial 
dwarf sequence from \cite{olsen84} which is well matched by our 
stars\footnote{Fiducials for different metallicities have recently been 
obtained by \cite{arna10}; here we use that of \cite{olsen84} for the sake of 
comparing directly with the original system.}. Because of the relatively 
bright magnitude range we sample, most of the late type stars are actually 
giants (cf.~left panels in Figure \ref{f:red}, where most of the stars 
in the bright magnitude range encompassed by dotted lines are in fact 
giants). Our sample 
is spread across several degrees in Galactic latitude, where reddening varies 
from a few hundredths to over a tenth of a magnitude (see Section 
\ref{sec:red}). In the central panel we thus use the virtually reddening--free 
indices $[c_1]=c_1-0.2(b-y)$ and $[m_1]=m_1+0.3(b-y)$ 
\citep[e.g.,][]{stro66,c75,olsen84}. It is clear that cool dwarfs are 
separated from giants and that there are a few cool M giants in our sample. 
For reference, 
in the right panel we also plot all dwarfs and subgiants from the GCS 
(which were observed by Olsen), a number of Red Giant Branch 
(RGB) stars from globular clusters of different metallicities 
\citep[][all standardized to the Olsen system]{gru98,gru00,gru02}, and 
our own observations of the open cluster NGC\,6819 in the {\it Kepler} 
field. Again, dwarf stars follow the Olsen's fiducial, while giants 
define upper sequences according to their metallicities.

\section{Determination of stellar parameters for {\it Kepler} 
targets}\label{sec:sp}

The purpose of the SAGA survey is to uniformly observe all stars across the 
{\it Kepler} field down to $V$ of $15^{\rm{th}}$ to $16^{\rm{th}}$ magnitude, 
independently of whether or not they have seismic information. We cross-match 
all stars with measured 
Str\"omgren photometry from the previous Section with all seismic 
targets in the {\it Kepler} field of view, i.e.~$\simeq 15,000$ giants from the 
{\it Kepler} Asteroseismic Science Consortium (KASC) \citep[][]{stello13,h14} 
and over 500 dwarfs \citep{chap13}. We detect 
about $95$\% of the seismic targets falling in the region sampled by SAGA so 
far, totaling $1,010$ targets ($29$ dwarfs and $981$ giants).
Further cross-matching of the sample with optical and infrared broad-band 
photometry (for the sake of the IRFM, see Section \ref{sec:teff}) and seismic 
analysis reduce the sample for which we determine full parameters to 989 
stars, implying a completeness of $93$\% for our final catalogue.
Most of the stars in our sample are giants (see also Figure \ref{f:fs}), but 
the procedure we adopt for determining parameters is the same whether a star 
is a dwarf or a giant (apart from the metallicity calibration, see Section 
\ref{sec:feh}). 

Multi-band photometry of stars in the {\it Kepler} field allows us to choose 
the most appropriate set of filters for each of the stellar parameters we wish 
to determine. One of the key advantages of our approach is that we use seismic 
information to improve upon the determination of photometric quantities. All 
parameters described in the following are derived iteratively and in a fully 
self-consistent manner, following the procedure first presented in 
\citet{vsa11,vsa12} and briefly summarized here. 
Details on each step are described in the relevant subsections.

Broad-band optical and infrared photometry is used to obtain effective 
temperatures ($\teff$), bolometric fluxes ($\fbol$) and angular diameters 
($\theta$) of each star via the IRFM (Section \ref{sec:teff}). This method 
depends weakly on the metallicity and surface gravity of a star, rendering it
ideal for breaking the degeneracy among these parameters, which instead 
seriously affects spectroscopic methods \citep[e.g.,][]{alonso96:irfm,c06}. 
However, being a photometric technique, reddening must be properly taken into 
account. 
For each star in our sample, we first take the color excess $\ebv$ and 
$\logg$ from the Kepler Input Catalogue \citep{kic}, while keeping 
metallicity fixed at $\feh=-0.2$, a value assumed to be roughly 
representative of the {\it Kepler} field \citep[e.g.,][]{vsa11,pi12,chap13}. 
These parameters are used as a starting point, but all our solutions converge 
independently of the initial choice. The effective temperature is derived 
using the IRFM with these initial $\logg$, $\feh$ and $\ebv$. This $\teff$ is 
then fed into an asteroseismic Bayesian scheme to obtain the mass and the 
radius of each star (Section \ref{sec:astero}).
Scaling $\theta$ with the asteroseismic radius, we compute the distance 
\citep{vsa12}. Using empirically calibrated, three dimensional Galactic 
extinction models (Section \ref{sec:red}) this distance is used to derive a 
new $\ebv$, which is then used to deredden the $uvby$ photometry and derive 
Str\"omgren metallicities (Section \ref{sec:feh}).
The above series of steps defines one iteration, and the updated set of $\ebv, 
\logg$ and $\feh$ determined at its completion is fed into the IRFM anew, 
and the procedure outlined above repeated until convergence is reached in all 
parameters: this was achieved within three iterations.

We note that in addition to our 
Str\"omgren photometry (used exclusively for deriving $\feh$), three broad-band 
systems are used in the IRFM: this implies three slightly different sets of 
$\teff$, $\fbol$ and $\theta$. Therefore, three sets of seismic parameters,
distances and thus reddening estimates (and $\feh$) are obtained even if the 
same reddening map is used. Within a given map, all these differences are 
relatively small.

Altogether we explore the use of three different broad-band photometric 
systems and three reddening maps for a total of 9 different self-consistent 
combinations of parameters. The effect of adopting different stellar models and 
pipelines for the seismic analysis has also been investigated. By comparing 
these values, we are able to obtain realistic error bars (Section 
\ref{sec:err}) as well as robust final parameters (Section \ref{sec:adopted}). 
A flow chart illustrating the entire procedure is shown in Figure \ref{f:flow}; 
in the following subsections we describe each step in detail, focusing first 
on the derivation of classical parameters and turning then to the asteroseismic 
analysis. 

We next discuss the rationale behind our choice of using different 
approaches to determine different stellar parameters, instead of for example 
implementing a global minimization scheme to match all photometric and 
seismic observables simultaneously. As described in the relevant subsections, 
effective temperatures, angular diameters, metallicities and reddening are 
determined using well-established and almost entirely empirical techniques. 
These empirically determined quantities are then fed into a Bayesian scheme 
built upon a grid of theoretical models. Extending this Bayesian scheme to 
include observed photometric indices, would make the determination of classical 
stellar parameters entirely dependent on 
theoretical models. In addition, it would carry uncertainties related on the 
standardization of synthetic colors (which are usually small) as well as on 
model flux deficiencies and/or uncertainties. The latter can be non-negligible 
in intermediate band filters, as well as towards the blue and ultraviolet 
wavelengths \citep{cvd14}. In our case, the dependence on synthetic flux is 
minimized by the multi broad-band photometry, as well as by the use of 
fully empirical calibrations in the intermediate band Str\"omgren system.

\subsection{Effective Temperatures and angular diameters}\label{sec:teff}

Effective temperatures are derived using the IRFM described in \cite{c10}.
This technique uses multi-band optical and infrared photometry to 
recover the bolometric and infrared flux of each star, from which its $\teff$ 
and $\theta$ can be computed. In our original formulation, 
implementing either 
Tycho2--2MASS or Johnson-(Cousins)--2MASS photometry, the zero point of the 
effective temperature scale is secured via solar twins, and the reliability of 
the angular diameters checked against dwarf and subgiant stars with 
interferometric measurements. 
This technique has already been used in a number of {\it Kepler} investigations 
\citep[e.g.,][]{vsa11,vsa12,hub12,chap13} and its accuracy 
now extended and validated also for giants \citep{c13}.

Here we implement the IRFM with three different optical systems (KIC $griz$, 
Johnson $BV$ and Sloan $g'r'i'$), while in all
instances 2MASS photometry is used in the infrared. While we defer more details 
on the quality of 2MASS photometry to Section \ref{sec:bin}, here it suffices 
to recall its excellent accuracy, with both median and mean errors of only 
$0.02$~mag in all three infrared bands. The implementation of the IRFM in the 
KIC $griz$--2MASS system is anchored to Tycho2--2MASS, requiring that on 
average the parameters derived in the two cases agree.
The KIC magnitudes are affected by 
color dependent zero-point offsets which are corrected according 
to \cite{pi12}. Fortunately, the AAVSO Photometric All-Sky Survey (APASS) 
offers an independent source of photometry to check the soundness of the 
results obtained using the KIC corrected photometry. We use the latest APASS 
Data Release (DR7) which covers 97\% of the sky in the magnitude range 
10 to 17 in the Johnson $BV$ and Sloan $g'r'i'$ filters. 
The advantage of implementing three different optical combinations in the IRFM 
is that the results in the Johnson--2MASS system are well standardized 
and much tested \citep{c10,c12}, while Sloan--2MASS allows us to check the 
consistency of the results obtained with the corrected KIC photometry plus 
2MASS. The 
comparison among the three systems also helps us to asses the reliability 
of the results and the overall error budget (see also Section \ref{sec:err} 
and Figure \ref{f:upj}). We stress here that optical broad-band photometry is 
preferable to Str\"omgren for the IRFM, since comparably little flux is 
encompassed within the intermediate band-width of the Str\"omgren system (which 
is anyway almost entirely covered by our broad-band photometry). Furthermore, 
Str\"omgren bands are designed to be more sensitive to both $\logg$ and 
$\feh$, which would make the IRFM estimates more dependent upon those 
parameters.

\subsection{Metallicities}\label{sec:feh}

Str\"omgren photometry is designed to effectively distinguish between dwarfs 
and giants and to provide reliable metallicity estimates in late type stars 
\citep[e.g.,][and references therein]{onehag09,arna10}. Much of the past 
literature has focused on metallicity calibrations for dwarf stars 
\citep[e.g.,][]{olsen84,schuster89,haywood02,nordstrom04,twarog07,c11} while 
considerably less attention has been devoted to giants. 
Most of the studies involving giants focused on the metal-poor regime 
\citep[e.g][]{faria07,cala07,adenthesis} and the only 
calibrations extending up to the solar metallicity are essentially those of 
\cite{gr92} and \cite{hi00}.

We have already derived a metallicity calibration for dwarfs and subgiants in 
\cite{c11} and here we extend it to giants. All calibrating stars in the 
metal poor regime have Str\"omgren photometry from the extensive work of 
\cite{gru98,gru00,gru02}, from which we cross-match stars with measured 
$\feh$ in the following monometallic\footnote{For M92, \cite{rs11} found that 
red giants are chemically homogeneous at the level of $0.07-0.16$~dex, 
although large non-homogeneity in neutron capture elements are disputed 
\citep{cohen11}.} globular clusters: 
NGC\,6341 \citep[M92, $\feh=-2.64$][]{rs11}, 
NGC\,6397 \citep[$\feh=-2.10$][]{lind11},
NGC\,6205 \citep[M13, $\feh=-1.58$][]{sne04},
NGC\,288 \citep[$\feh=-1.22$][]{carre09},
NGC\,6838 \citep[M71, $\feh=-0.82$][]{carre09},
NGC\,104 \citep[47~Tuc, $\feh=-0.75$][]{carre09}. In the metal-rich regime we 
use our own Str\"omgren photometry, with single member, seismic giants in the 
solar metallicity open cluster NGC\,6819 \citep{stello11}, plus seven field 
giants we have in common with the 
spectroscopic sample of \cite{thy12}. Altogether our Str\"omgren metallicity 
sample of giants comprises $199$ stars. We take reddening values for stars in 
globular clusters from the latest version of the \cite{ha96} 
catalogue\footnote{http://www.physics.mcmaster.ca/Globular.html}. This 
provides robust estimates, given that the majority of them has very low $\ebv$, 
and the absence of any substantial differential reddening \citep[even for 
NGC\,6397 and NGC\,6838 which suffer from somewhat higher extinction, 
see][]{mi12}. For NGC\,6819 we adopted 
$\ebv = 0.14$ while for the remaining field giants reddening is estimated in 
much the same way as for our other {\it Kepler} targets (see Section 
\ref{sec:red}).

We tested different functional forms relating Str\"omgren colors\footnote{In 
the following we will refer to $c_1$, $m_1$ and $b-y$ with the implicit 
understanding that when deriving metallicities these indices must first be 
dereddened.} to $\feh$. Instead of including all possible combinations of 
indices in some high-order polynomial, we started with a simple multi-linear 
dependence on the three Str\"omgren indices, and introduced mixed and higher 
order terms after verifying that they improved the residuals. The form 
adopted is the following:
\begin{displaymath}
\feh \,=\,- 9.037 (b-y) \,+\, 4.875 c_1 \,+\, 17.187 m_1 \,+\, 3.672 (b-y) c_1 
      \,-\, 9.430 (b-y) m_1  
\end{displaymath}
\begin{equation}\label{eq:meca}
 -\, 12.303 c_1^2 \,-\, 17.405 m_1^2 \,+\, 6.159 (b-y)^2 \,+\, 12.917 m_1^3 
\,-\, 0.972,
\end{equation}
which applies to giants with $0.45 \leq (b-y) \leq 1.10$, 
$0.08 \leq m_1 \leq 0.92$, $0.08 \leq c_1 \leq 0.59$ and 
$-2.6 \lesssim \feh \lesssim 0.5$~dex, with a standard 
deviation of $0.15$~dex (Figure \ref{f:mecal}). Notice that we apply this 
calibration to all our giants (cf.~Figure \ref{f:fs}), while for dwarfs we 
rely on the similar calibration derived in \cite{c11}. As there are no 
calibration giants above $\feh \simeq 0.5$~dex in the sample, if a star 
has a metallicity higher than this value, or if its Str\"omgren indices fall 
outside of the applicability range determined above, we flag the measurement. 
This and the $uvby$ photometric quality flag determined in Section 
\ref{sec:sspc} identify stars with the most reliable metallicities based on 
the Str\"omgren photometry. As a result, these flags help to diagnose 
cases where stars with slightly unusual colors might return unrealistic 
metallicities.

Equation \ref{eq:meca} differs from the functional form often 
used for giants, in particular it includes a dependence on the $c_1$ 
index, which was not measured in many past works (the low flux in the 
$u$ band of relatively faint giants being too time consuming to measure with 
photoelectric photometers). The standard deviation of our giant calibration is 
somewhat larger than for dwarfs \citep[$\lesssim 0.1$~dex,][]{c11}; this could 
partly originate from the fact that the calibration sample already carries
some uncertainty in the adopted reddening values. As we see later, the typical 
reddening uncertainty is around $\sigma_{E(B-V)}=0.02$, 
and this affects Str\"omgren metallicities by $0.09$~dex on 
average. We thus take a conservative approach and quadratically sum this error 
to the above standard deviation, yielding a typical uncertainty of $0.17$~dex 
for our metallicities. 
As the SAGA survey continues, we plan to improve and expand the above 
calibration with a 
larger and more uniform sample of giants having spectroscopic metallicities 
and abundances. In addition, Str\"omgren indices are known to be sensitive to 
more than just $\feh$ in cool stars and we plan to investigate this effect as 
well \citep[e.g.,][]{yong08,c11,carre11}. 

The open cluster NGC\,6819 offers the possibility to check the consistency 
between the metallicity calibration derived here for giants, and that used 
for dwarfs in our previous analysis of the solar 
neighborhood \citep{c11}, as well as for the few dwarfs present in this 
sample (cf.~Figure \ref{f:fs}). We identify dwarfs and giants in 
NGC\,6819 by using the same procedure for both: we cross-match our 
Str\"omgren observations with the radial velocity catalogue 
of \cite{hole09}, and retain only single-stars with cluster membership 
probability $>80$\% and within $\simeq 7$~arcmin from the centre of the 
cluster. Notice that with this procedure, some of the cluster giants are not 
the same used for our metallicity calibration, which had more secure seismic 
membership \citep{stello11}. The comparison in Figure \ref{f:cmd} shows 
that metallicities of dwarfs and giants agree on average. However, two things 
should be noticed. Firstly, the dwarfs exhibit a larger scatter, which is 
partly due to the larger photometric errors at fainter magnitudes, and because 
the sensitivity of Str\"omgren indices to metallicity is reduced for warmer 
dwarfs than for giants. Secondly, the peak of the two metallicity 
distributions differs by some $0.08$~dex, with the giants being more metal-poor.
Such an offset is not surprising, given that there might be 
a difference in the effective temperatures underlying the spectroscopic 
measurements used to build the dwarf \citep{c11} and the 
giant (derived here) metallicity calibrations. In fact, 
spectroscopic $\feh$ in giants are often derived using different 
flavors of the \cite{alonso99} $\teff$ scale. 
This is about $100$~K cooler than the one now preferred for dwarfs and 
giants \citep{c10,c13}. Since the metallicity calibration for dwarfs used 
spectroscopic metallicities based on a $\teff$ scale hotter 
than \cite{alonso99}, the aforementioned $100$~K difference in giants can 
easily account for the metallicity offset found here.
Obviously, this zero point difference can be corrected, and for 
convenience the last term in Equation \ref{eq:meca} is already shifted by this 
amount (a later comparison done in Figure \ref{f:6819} confirms that with such 
a correction giants in the cluster have indeed solar metallicity).

One could argue that $\feh$ should be slightly {\it lower} in dwarfs 
than giants belonging to the same stellar population due to the settling of 
heavy elements in main-sequence stars \citep[e.g.,][]{korn07}. However, because 
of the uncertainties discussed above and the relatively young age of the 
cluster \citep[$\simeq2.5$~Gyr, e.g.,][]{yang13,jef13,sand13}, we believe this 
effect to be well within the uncertainties of our photometric metallicities.

\subsection{Reddening}\label{sec:red}

\begin{figure}
\plotone{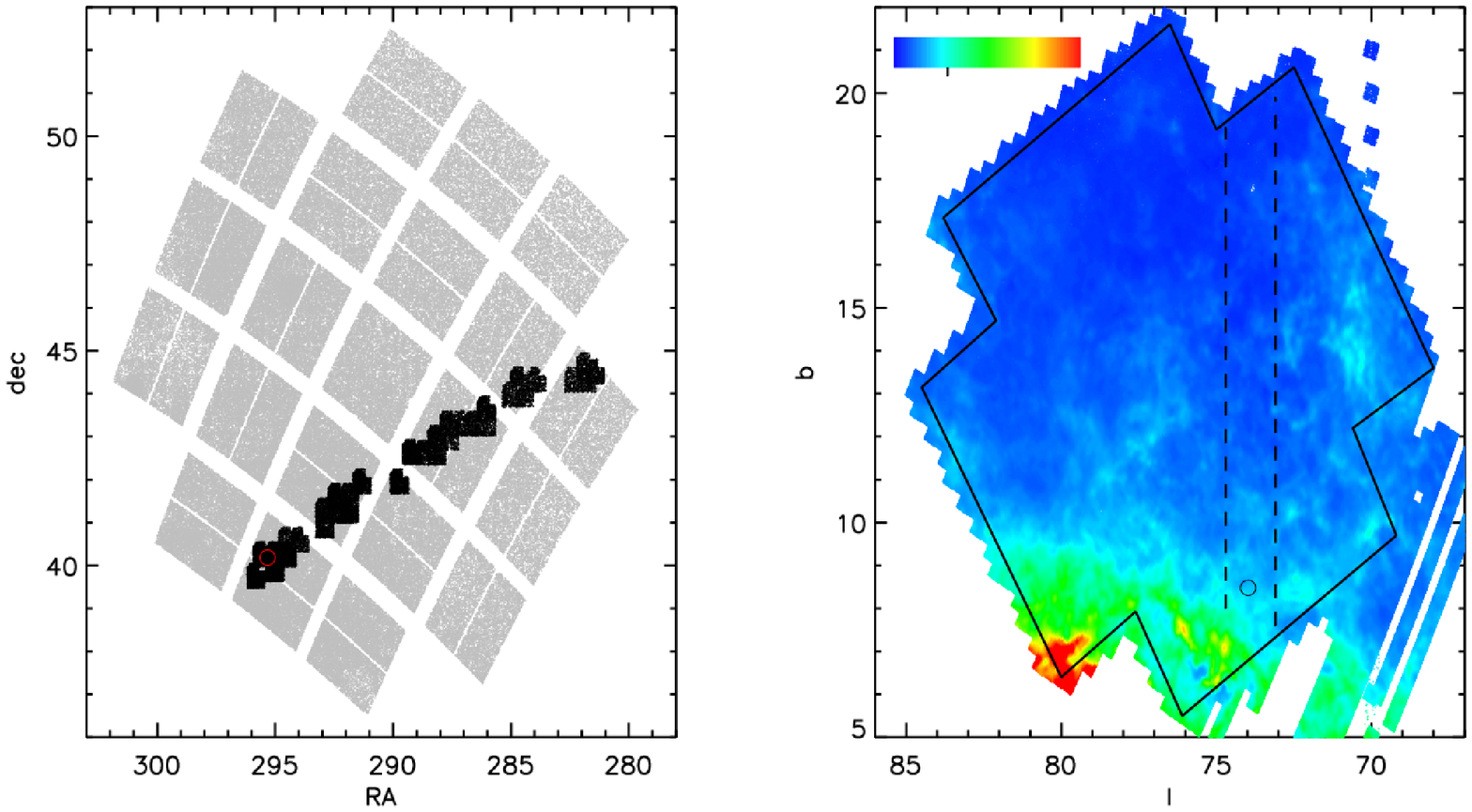}
\caption{{\it Left panel:} The {\it Kepler} field as seen via the 42 CCDs on 
board of 
the spacecraft. Overplotted in black is the full sample of stars with $uvby$ 
magnitudes measured during the first run of the SAGA survey. The circle 
marks the position of the open cluster NGC\,6819. {\it Right panel:} {\it 
Kepler} 
field (defined with continuous black lines) plotted in Galactic latitude ($b$) 
and longitude ($l$) as seen in the \cite{sfd98} map. Linear color code 
corresponds reddening in the range $0<\ebv<1$. The tick mark in 
the upper bar 
identifies $\ebv=0.3$ which is the highest value in the \cite{sfd98} map over 
our observed stripe. Vertical dashed lines mark the observed stripe, and the 
open circle indicates the position of NGC\,6819.\label{f:fov}}
\end{figure}

\begin{figure}
\epsscale{1}
\plotone{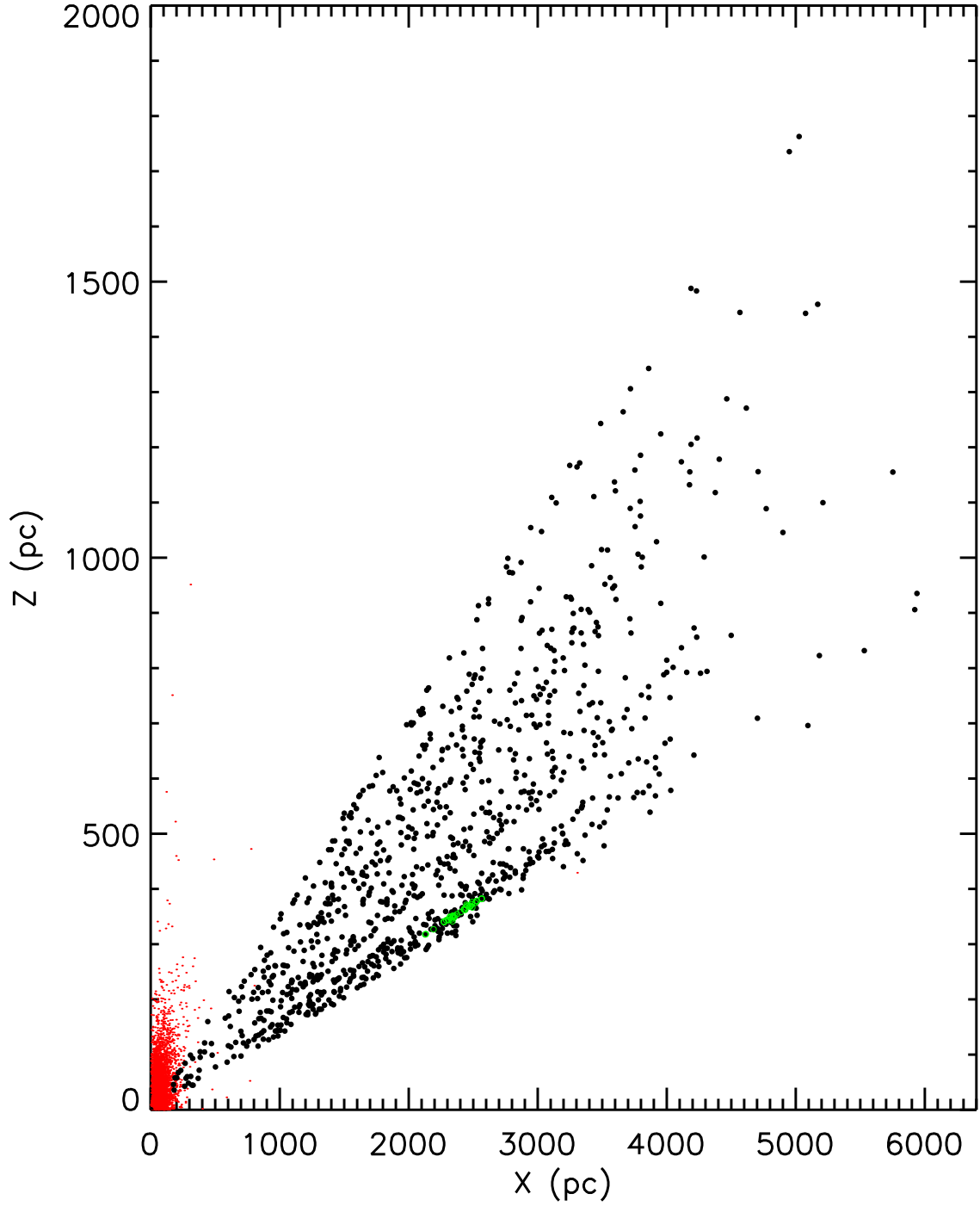}
\caption{Distance distribution for all seismic targets currently in SAGA. 
Overplotted in red is the volume surveyed by the Geneva-Copenhagen Survey. 
Highlighted with green open circles are giants belonging to NGC\,6819. 
Distances have been derived as described in Section \ref{sec:sp}.
\label{f:trigo}}
\end{figure}

\begin{figure}
\plotone{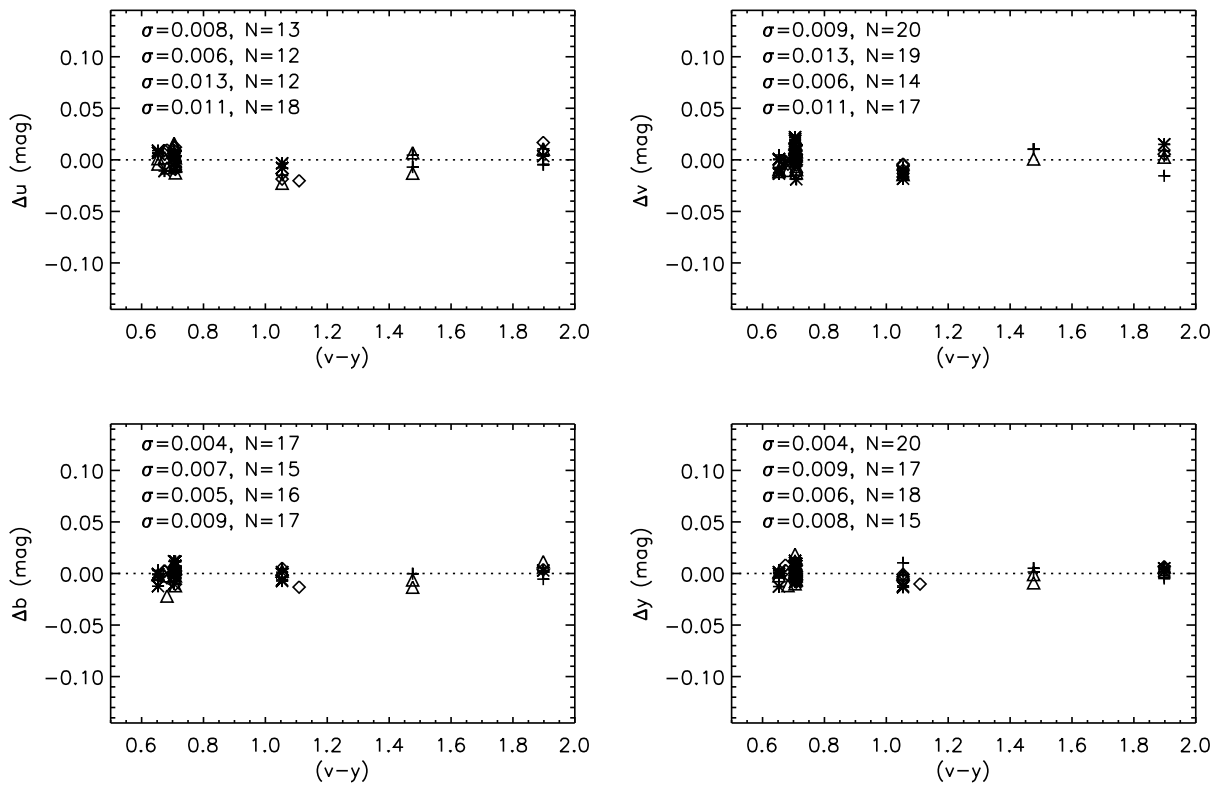}
\caption{Residuals of the standard stars (calibrated minus standard) as 
function of their $(v-y)$ color. The number of points ($N$) and the scatter 
($\sigma$) for each photometric night (different symbol for each night) is 
indicated.
\label{f:standards}}
\end{figure}

\begin{figure}
\plotone{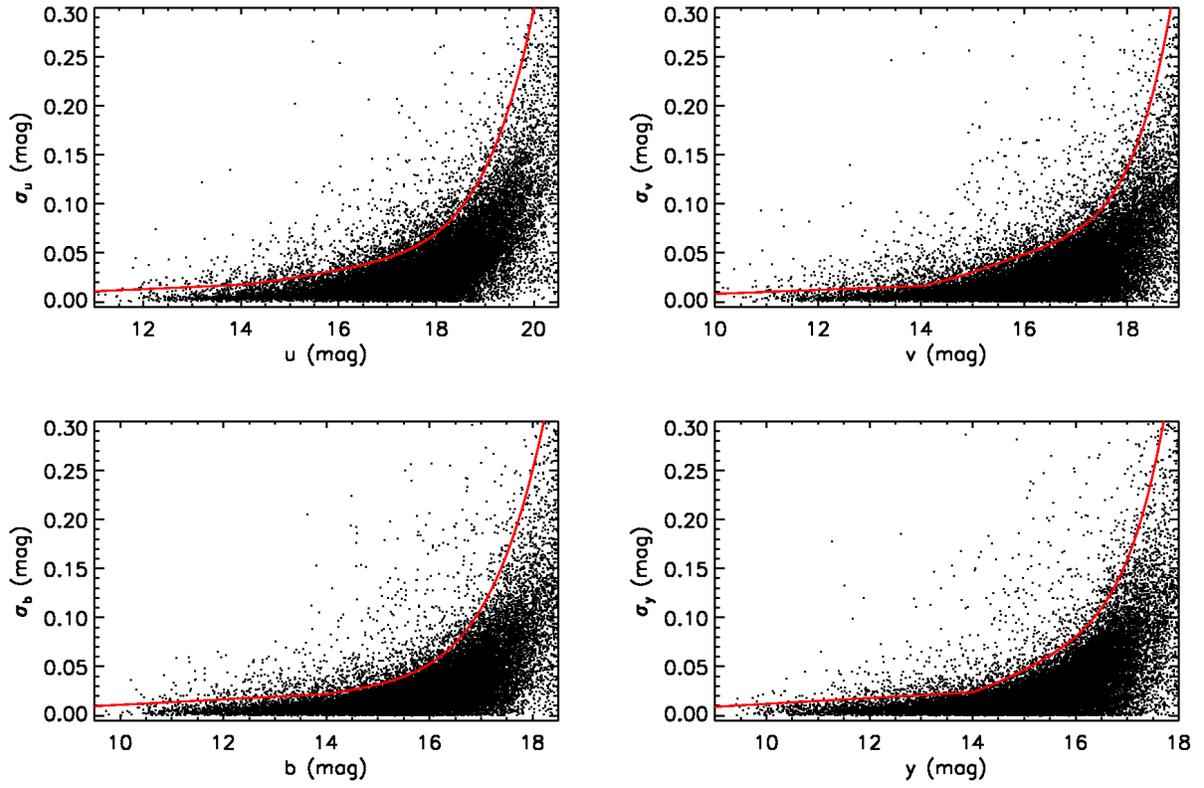}
\caption{Final errors in $u,v,b$ and $y$ photometry. Stars above the 
continuous ridge-lines are labelled in the analysis as explained 
in Section \ref{sec:sspc}.\label{f:errors}}
\end{figure}

\begin{figure}
\plotone{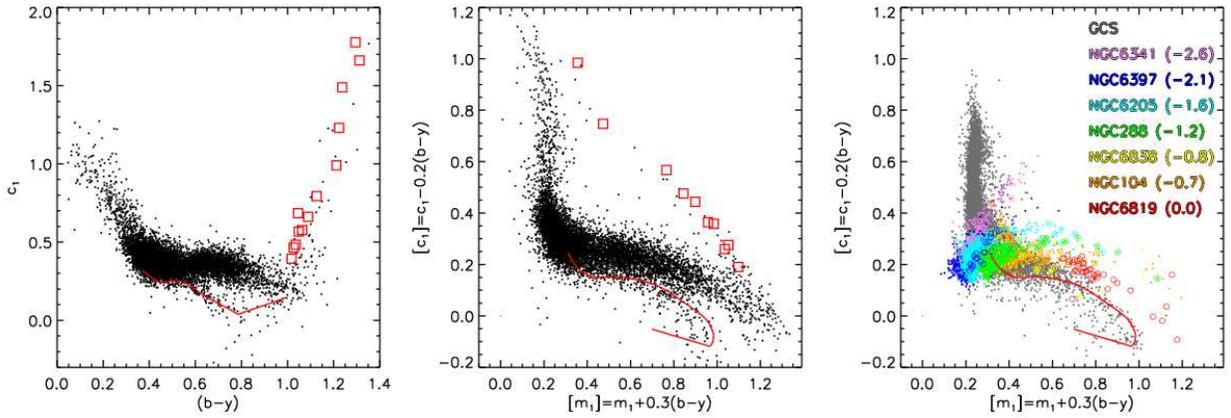}
\caption{{\it Left panel:} $c_1$ vs.~$(b-y)$ diagram for stars in our sample 
with $y < 15$~mag and photometric errors below ridge-lines of Figure 
\ref{f:errors}. Open squares are cool M giants, while the continuous line 
represents the dwarf sequence, both from \cite{olsen84}. {\it Central panel:} 
same as 
previous, but for reddening-free indices $[c_1]$ vs.~$[m_1]$. 
{\it Right panel:} reddening-free indices for dwarfs and subgiants from the 
GCS as well as for a number of RGB stars in stellar clusters (metallicities in 
parenthesis), all standardized to the system of Olsen. Open circles are stars 
with spectroscopic $\feh$ used for our giants metallicity calibration.\label{f:strom}}
\end{figure}

\begin{figure}
\epsscale{.90}
\plotone{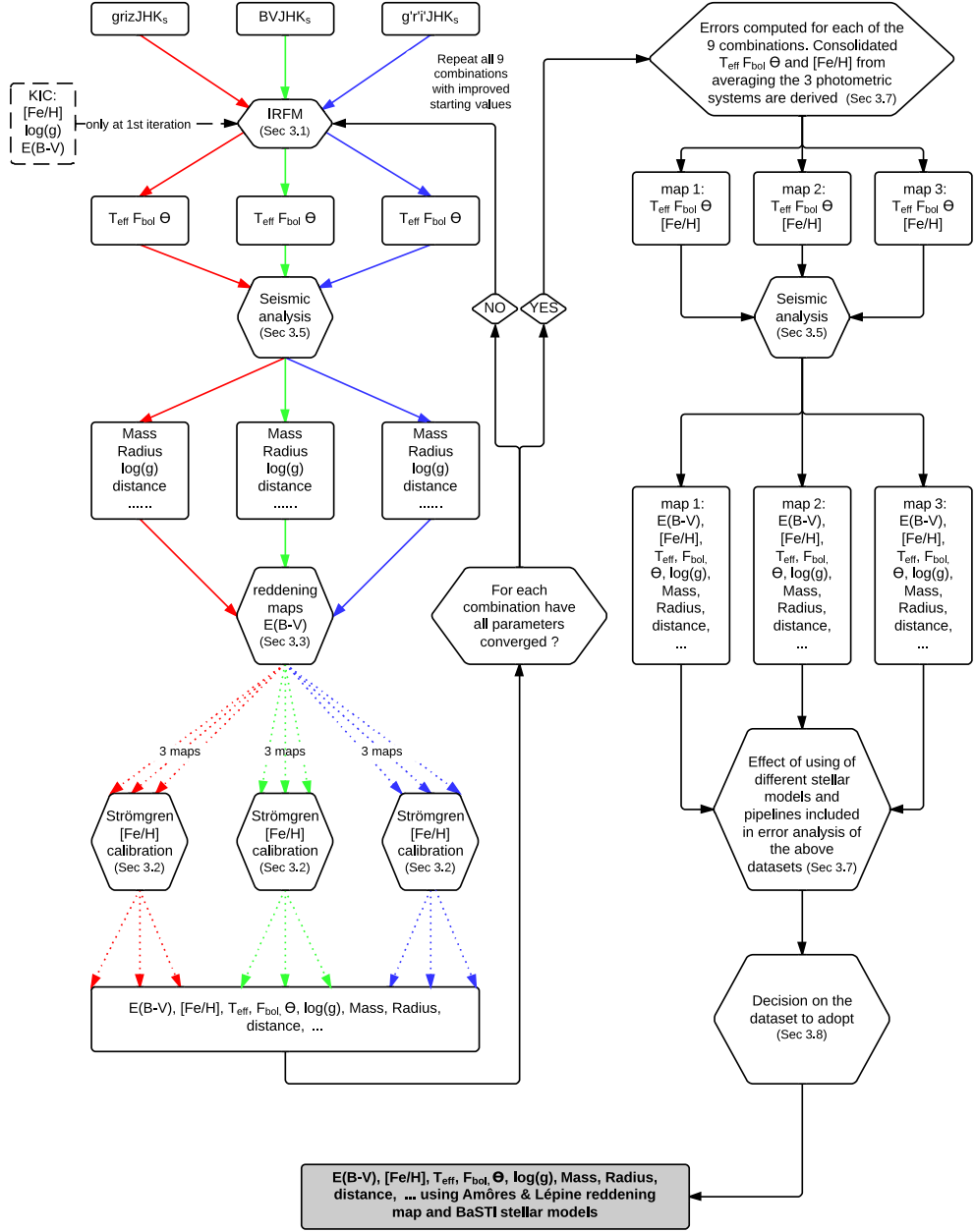}
\caption{Flow chart of the procedure adopted to intertwine the derivation of 
classical and seismic parameters. Reddening maps 1, 2, and 3 are the 
\cite{drimmel03} and the two flavors of the \cite{al05}.\label{f:flow}}
\end{figure}

\begin{figure}
\epsscale{.70}
\plotone{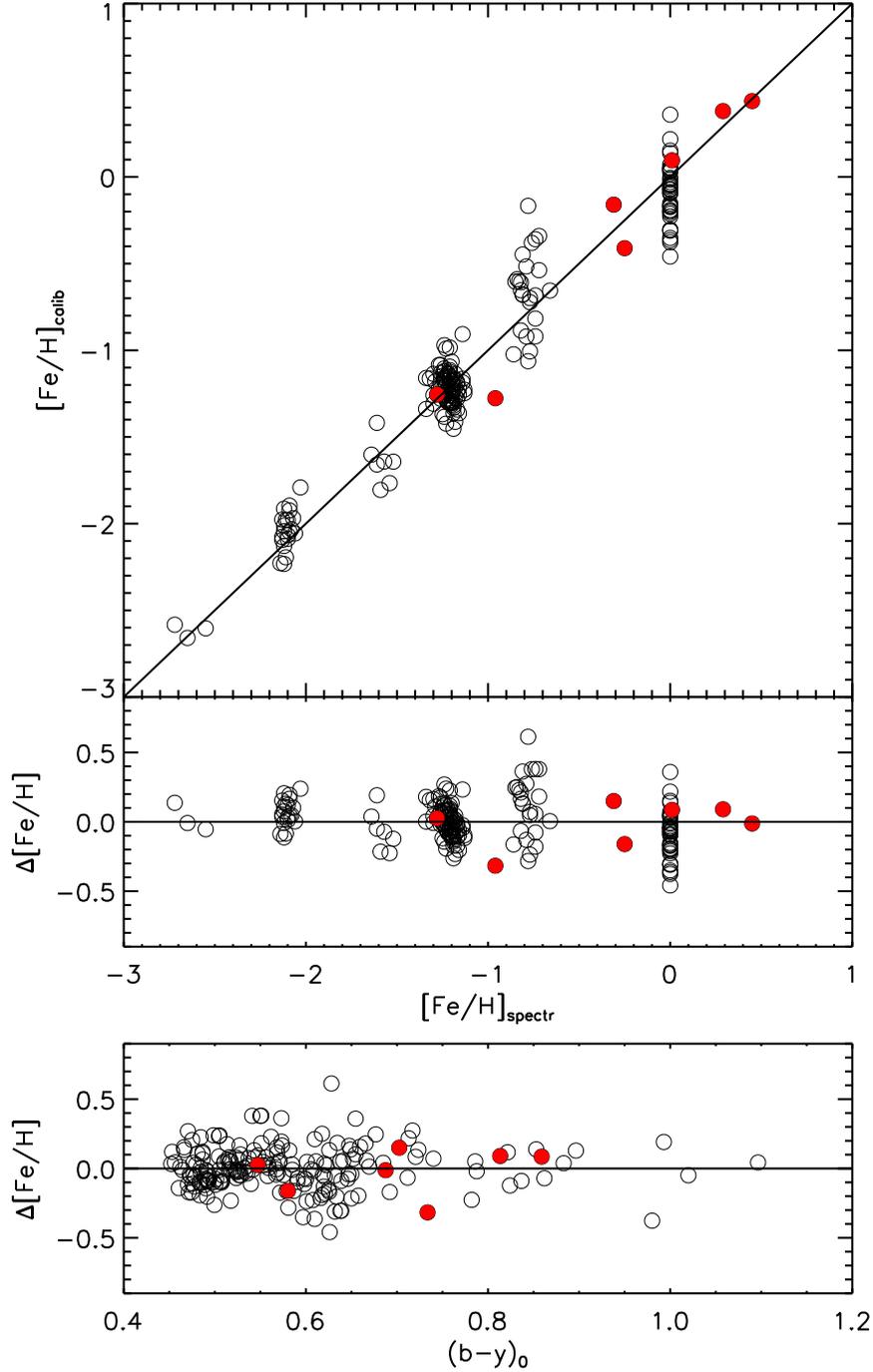}
\caption{{\it Top panel:} spectroscopic versus photometric metallicities 
obtained using the calibration presented in Section \ref{sec:feh} for giants. 
Open circles are giants in clusters, while filled circles are field giants 
from \cite{thy12}. 
{\it Middle panel:} same as above, but showing residuals (photometric minus 
spectroscopic). {\it Bottom panel:} same as in previous one, but as function 
of dereddened $(b-y)$, which is a proxy of the effective 
temperature.\label{f:mecal}}
\end{figure}

\begin{figure}
\epsscale{.99}
\plotone{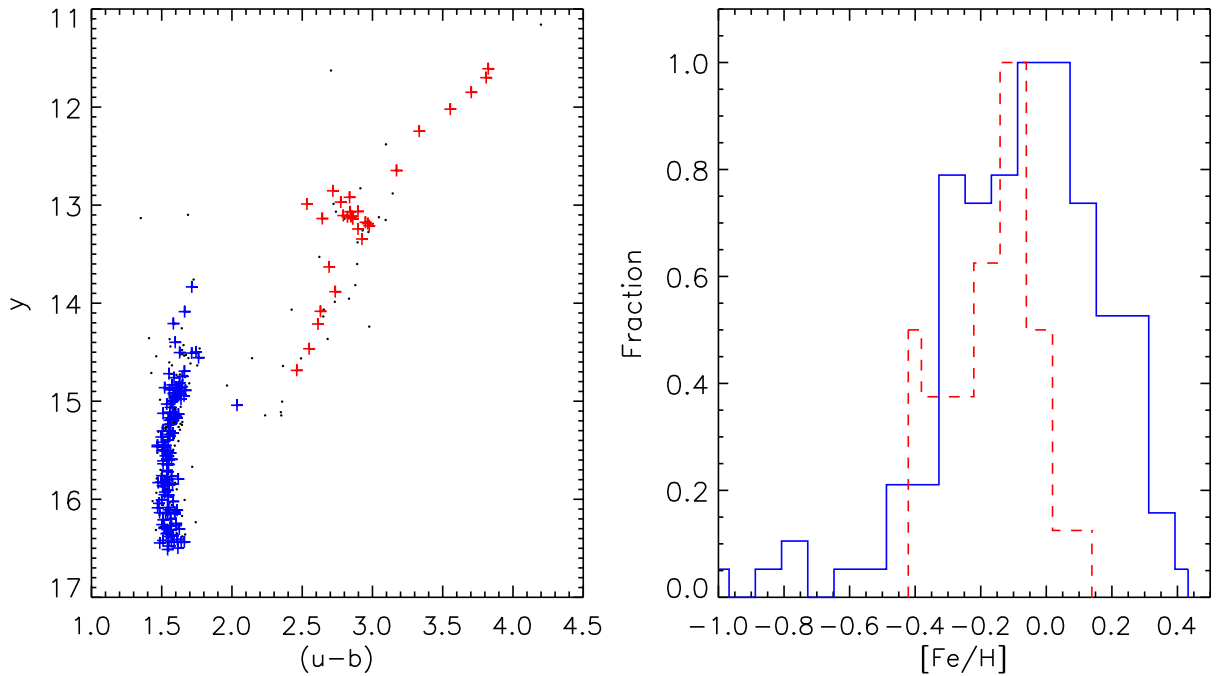}
\caption{{\it Left panel:} NGC\,6819 color-magnitude diagram. Only single-stars 
having radial velocity membership probabilities higher than $80$\% in 
\cite{hole09} are shown. Highlighted in 
red (blue) are giant (main-sequence) stars within $\simeq 7$~arcmin from the 
cluster centre and with photometric errors below the ridge-lines of 
Figure \ref{f:errors}. {\it Right panel:} Str\"omgren metallicity histogram 
for the same dwarfs (continuous) and giants (dashed line) highlighted in the 
left panel. The difference between the peaks of the metallicity distributions 
of dwarfs and giants ($0.08$~dex) has been corrected as explained in Section 
\ref{sec:feh}.\label{f:cmd}}
\end{figure}

Even though the Galactic stripe observed for this work avoids excessively high 
values of color excess, non-negligible reddening is still present and must be 
corrected for before we can use photometry to derive stellar parameters. 
We did a comparison between the color excess reported in the Kepler Input 
Catalog for over 13 million objects, and $\ebv$ derived at their same position 
in the sky from the \cite{sfd98} map (Figure \ref{f:fov}). This revealed that 
that extinction in the Kepler Input Catalog lacks fine 
structure. This is not surprising since reddening in the Kepler Input Catalog 
is estimated using a simple dust model which assumes a smooth exponential disk 
with a scale height of $150$~pc and 1~mag of attenuation in $V$ band per kpc in 
the plane. Thus, color excess from the Kepler Input Catalog is used only as a 
starting point in our iterative scheme (Section \ref{sec:sp}).

Because we derive distances for all our stars, we can 
use three-dimensional extinction models. We use two different ones; that of 
\cite{drimmel03} and \cite{al05}.
Drimmel's map is based on a dust distribution model applied to three 
Galactic density components and rescaled to recover the observed far-infrared 
emission as observed by the COBE satellite. The \cite{al05} map is 
based on the measured gas density distribution, and translated into extinction 
under the assumption that gas and dust are well mixed. This model is available 
for two configurations, one assuming that the Galaxy is axisymmetric and one 
taking into account the spiral structure of the Galaxy. 

The scale of the geometrical components used in all these models of the Galaxy 
is poorly known, and therefore the spatial distribution of extinction is 
uncertain. In fact, all these models suggest that rescaling factors can 
be used, should these be needed to satisfy reddening constraints in 
particular lines-of-sight.

Once again the open cluster NGC\,6819 offers an important calibration point. It 
is located at the base of our Galactic stripe (Figure \ref{f:fov}) and its 
giants are in the same 
magnitude range of the other {\it Kepler} giants with detected oscillations
(Figure \ref{f:cmd}), thus broadly representing the typical mean distance 
probed in our investigation (see also Figure \ref{f:trigo}).
Its reddening is also well constrained: we adopt $\ebv=0.14$ from \cite{b01}, 
which falls within the $0.12-0.16$ range of values found using 
different methods \citep[e.g.,][]{rv98,jef13}. 
Using our Str\"omgren photometry
we apply the empirical method described in \cite{milone12} to derive a 
$14 \times 14$~arcmin map of differential reddening centred on the cluster, 
which shows spatial variations of order $\pm 0.02$ mag. This is consistent 
with the result of \cite{plata13}, and reassures us
that even at the lowest latitudes of our sample, differential 
reddening --whilst present-- is not a major concern for our analysis. 

We thus rescale the \cite{drimmel03} and \cite{al05} models to have 
$\ebv=0.14$ at the position and distance of NGC\,6819 \citep[$2.4$~kpc from 
isochrone fitting,][]{yang13}, and use these 
recalibrated maps to estimate reddening for all our stars. Incidentally we 
notice that after 
this rescaling, the axisymmetric and the spiral model of \cite{al05} are 
virtually identical (always to better than $0.01$~mag) over the Galactic 
coordinates and distances covered in this investigation. For consistency we 
perform our analysis using the three models, but in practice 
we will always refer to both \cite{drimmel03} or \cite{al05} with the 
implicit assumption that the two flavors of the latter have been 
compared and that no substantial difference is found in the parameters derived. 

\begin{figure}
\epsscale{.72}
\plotone{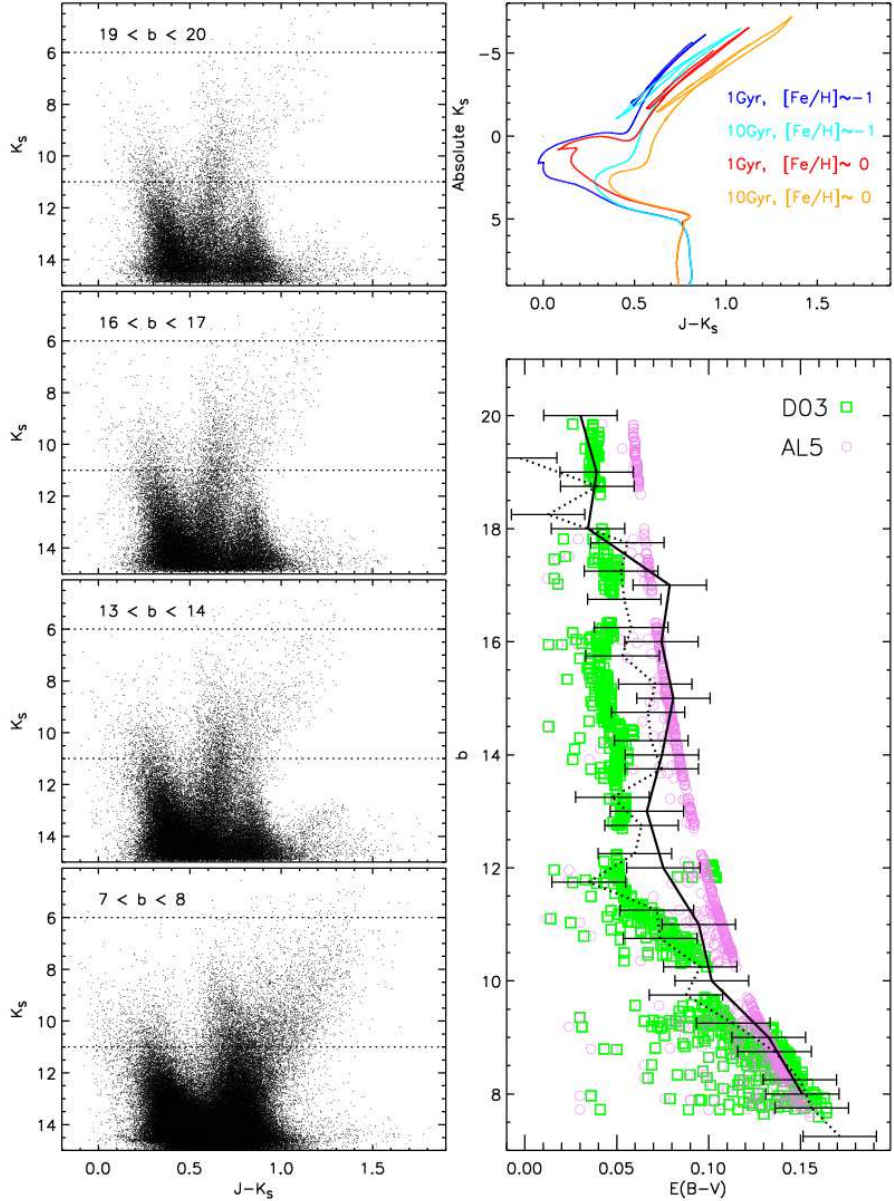}
\caption{{\it Left panels:} observed $J-K_S$ vs. $K_S$ diagram for stars with 
good 2MASS photometry at selected intervals in Galactic latitude $b$, and with 
longitude $ | l - 72.5^{\circ} | < 2.5^{\circ}$.
Dotted horizontal lines indicate the $K_S$ magnitude range used to empirically 
derive reddening (see text for details). 
{\it Top right:} BaSTI isochrones in the 2MASS system for two values of 
metallicity and age \citep{pietrinferni04}. 
{\it 
Bottom right:} reddening derived using the procedure described in Section 
\ref{sec:red}, sampling the 2MASS color-apparent magnitude diagram each 
$0.5^{\circ}$ (dotted) or $1^{\circ}$ (continuous lines), after scaling to the 
$\ebv$ of NGC\,6819. Open circles and 
squares are reddening values derived for our {\it Kepler} stars using the 
\citealt{al05} (AL05) and \citealt{drimmel03} (D03) maps.
\label{f:red}}
\end{figure}

With respect to extinction models, 2MASS photometry offers an alternative and 
empirical way to gauge the reddening .
To this end, we select from the 2MASS Point Source Catalog \citep{skr06} a 
region roughly centred at the same longitude of our Galactic stripe, 
i.e.~$|l-72.5^{\circ}| <\pm 2.5^{\circ}$, and retain only targets having good 
infrared photometric flags\footnote{http://www.ipac.caltech.edu/2mass/releases/allsky/doc/sec2\_2a.html} (i.e.~quality ``AAA'' and blend ``111''). Figure \ref{f:red} 
shows the 
$J-K_S$ vs.~$K_S$ plane for sources at four different latitudes. At the 
highest latitudes, where reddening is lowest, three main features 
are obvious: 1) the bluemost overdensity, which mainly comprises main-sequence 
and turn-off stars, 2) the central one mostly made up of giants, and 3) the 
rightmost and fainter group of cool dwarfs. Moving to lower Galactic latitudes 
these 
features blur, and a prominent puff of stars appears at the redmost colors. 
Reddening is indeed the main factor in shaping these morphological 
changes across the panels, as both age and metallicity play a minor role in 
the near-infrared color-magnitude diagram. This is shown with isochrones in 
the top right panel of the same figure. The main position of the red 
giant population is always located at approximately the same color 
$J-K_S \simeq 0.6-0.7$ regardless of the age ($1$ and $10$~Gyr) and metallicity 
($\feh \simeq -1$ and $\sim 0$) of the underlying population. We note that the 
isochrones shown here span a range considerably larger than the mean variation 
measured in this part of the disk.
Thus, we can turn this argument around, to derive the value of reddening at 
each Galactic latitude such that the red giant population is restored to the 
same $J-K_S$. This can be used to verify the adopted Galaxy model reddening 
maps.

From this 2MASS stripe, we sample Galactic latitudes between 
$7^{\circ}$ 
and $20^{\circ}$ at constant steps of either $0.5^{\circ}$ or $1^{\circ}$ (to 
check different resolution effects) and each time derive the corresponding 
$J-K_S$ vs.~$K_S$ diagram. We consider only stars having apparent magnitudes 
$6< K_S < 11$. This choice is representative of the magnitudes, 
and 
hence distances, of the bulk of our {\it Kepler} giants, and it also minimizes 
contamination from dwarfs (the reddest overdensity of dwarfs in the left 
panels of Figure \ref{f:red} starts at somewhat fainter magnitudes). 
We then sample the reddening space in the range 
$0 \leq \ebv \leq 0.4$ in steps of $0.01$, each time dereddening the 
color-apparent magnitude diagram at the value of $\ebv$ considered. For the 
sake of our analysis each color-apparent magnitude diagram is reduced to a 
histogram of the relative number of stars as function of $J-K_S$ color.
The purpose is to derive a reddening map whose values are
differential with respect to a reference value. Here, the histogram at the 
highest latitude sampled is taken as reference, and hence its color magnitude 
diagram is uncorrected for reddening, even if it could have a non-zero value 
of $\ebv$.
All histograms derived at lower 
latitudes and for different values of $\ebv$ are benchmarked against this. 
Then, at each latitude we find the value of $\ebv$ which best matches 
the reference histogram in the color range $0.5 \leq J-K_S \leq 1.5$. These 
last values are set somewhat arbitrarily, to avoid the inclusion of the 
age-sensitive turn-off population and to probe the population of red giants 
well enough. We verified that changing these limits within reasonable values 
($\simeq \pm 0.1$~mag) does not severely affect our results. 
This differential approach can then be put on an absolute 
scale should the value of reddening be known in one of the fields: this is 
indeed the case at the location of the open cluster. All $\ebv$ 
solutions are thus shifted by the amount needed to pass through the reddening 
of NGC\,6819 at $b=8.5^{\circ}$. As already mentioned, we sample Galactic 
latitudes in steps of both $0.5^{\circ}$ and $1^{\circ}$ to check whether our 
results depend on the adopted binning: the agreement is quite reassuring, 
always within a few hundredths of a magnitude, although values of 
$\ebv$ derived sampling each $0.5^{\circ}$ appear to be more noisy. This is 
expected because of the lower number of stars when using smaller bins, 
especially towards the highest latitudes. From this scatter we estimate 
$\sigma_{E(B-V)} = 0.02$ on average for the population {\it as a whole}.
The reddening maps built with such a procedure have a rather coarse 
resolution, but benefit from being empirical and completely 
independent from both \cite{drimmel03} and \cite{al05}. 

The overall level of agreement between different maps and our empirical method 
is quite comforting provided that NGC\,6819 is used as anchor point, and gives 
us confidence that our parameters are not severely affected by reddening. 
Notice that all our stellar parameters are derived using both the 
\cite{drimmel03} and \cite{al05} maps, while the empirical method developed 
here does not contain any distance information on single stars. 
However, besides validating the adopted extinction maps, the empirical method 
allows us to estimate the average uncertainty in reddening, at the level of 
$0.02$~mag, and thus to assess the robustness of the stellar parameters we 
derive in the presence of this uncertainty.

\subsection{Global asteroseismic parameters and evolutionary phase classification}\label{sec:glob_astero}

Stellar oscillations driven by surface convection, such as those observed in 
the Sun and red giants, are visible in the power spectrum of time series 
photometry as a series of Lorentzian-shaped peaks whose peak height is 
modulated in frequency by a Gaussian-like envelope 
\citep[e.g.,][]{cm13}. Two quantities can be readily extracted from this 
oscillation pattern: the average separation between peaks of the same angular 
degree and consecutive radial order, the large frequency separation 
$\Delta\nu$, and the frequency of maximum amplitude $\nu_\mathrm{max}$ 
\citep[e.g.,][]{Ulrich:1986ge,brown91}. These 
global asteroseismic parameters have been determined for the full sample of 
giants that we consider here. A few dwarfs are also present in the 
sample, and their classification comes from \cite{chap13}.

Provided that the time resolution and quality of the observations are good 
enough, 
additional information can be extracted from the power spectrum of red giants 
thanks to the presence of {\it mixed modes} \citep[see e.g.,][for a 
review]{Bedding:2011wl}. These oscillations have an acoustic character in the 
outer layers while behaving like gravity waves in the deep interior of a star, 
providing unique information about the stellar core. Stars evolving along the 
red giant branch, burning hydrogen in a shell around the inert helium core, 
are very difficult to distinguish from their helium core burning red clump 
counterparts by classical photometric and spectroscopic observations. However, 
RGB stars have a radiative core, while clump giants have a convective one. 
Measurement of the average spacing between mixed dipole modes, clearly 
separates these stars into two distinct populations of giants 
\citep{Bedding:2011il,Mosser:2012jy}.

We have analyzed \textit{Kepler} long-cadence data \citep{jenkins10} through 
Quarter 15 for all giants in our sample. About 30\% of the sample has been 
observed continuously throughout the mission (with a maximum dataset length of 
1,350 days), while 70\% have more than 10 quarters of data available. 
Simple-aperture photometry (SAP) data was used for our analysis. Instrumental 
flux discontinuities were corrected by fitting linear functions to 5-10 day 
subsets before and after each discontinuity \citep[see, e.g.,][]{garcia11}. 
Finally, for all time series a quadratic Savitzky-Golay filter 
\citep{savitzky64} was applied to remove additional variability due to stellar 
activity and instrumental effects. Due to the wide range of oscillation 
periods covered in our sample we used two different filters: for stars with 
oscillation periods $>1$ day we applied a $50$~day filter, and for all 
remaining stars a $10$~day filter was used.

We have classified the red giant stars into two main groups: those burning 
only hydrogen in a shell located on the RGB and those burning helium in the 
core in the red clump (RC).  
To distinguish these two groups we looked for the seismic signature of the
stellar core properties according to \citet{Bedding:2011il}. Specifically, 
we measured the period spacing of the dipole modes following the approach by
\citet{stello13}.  We were able to measure the period spacing and unambiguously 
classify 427 stars in this way (169 RGB and 258 RC). The remaining 
giant stars did not show sufficient signal-to-noise in the frequency power 
spectra to provide reliable classification.

To extract the global oscillation parameters $\Delta\nu$ and $\nu_{\rm{max}}$ we 
have used the analysis pipeline described by \citet{huber09}, which has been 
shown to agree well with other methods 
\citep{hekker11,verner11}. Uncertainties for each parameter were determined 
through Monte-Carlo simulations, as described in \citet{Huber:2011be}. 
For each value we added in quadrature an uncertainty arising from the use of 
solar reference values to estimate stellar properties (Section 3.5). 
The final median uncertainties on $\Delta\nu$ and $\nu_{\rm{max}}$ were 0.7\% 
and 1.7\%, reflecting the exquisite signal-to-noise of the \textit{Kepler} data.

\subsection{Masses, radii and distances}\label{sec:astero}

We have used grids of isochrones in combination with a Bayesian analysis
that includes asteroseismic quantities to determine the stellar parameters of
our sample.
The reference isochrones are constructed from the non-canonical
BaSTI models, which include core overshooting during the H-core burning phase 
and do not consider mass-loss. These isochrones have been computed explicitly 
for this work, while the BaSTI grid publically 
available\footnote{www.oa-teramo.inaf.it/BASTI} assumes a 
\cite{rei75} mass-loss parameter of either $\eta=0.4$ or $\eta=0.2$.
The helium to metal enrichment ratio is fixed 
at $\Delta Y / \Delta Z=1.45$, consistent with the value inferred from the 
broadening of the low main-sequence \citep[][]{c07} and from models of stellar 
nucleosynthesis and chemical evolution of the Galactic disk 
\citep[e.g.][]{m92,cp08}. 
A detailed account of the input physics of the models is given in 
\citet{pietrinferni04}. We note that measurements of period spacings between 
mixed modes in clump stars 
observed by {\it Kepler} \citep{Mosser:2012jy} can only be reproduced if 
mixing beyond the formal Schwarzschild boundary of the He-burning convective 
core is included \citep{Montalban:2013gf}. In BaSTI models, this extra mixing 
is accounted for by including semiconvection during the He-core burning phase. 

The Bayesian analysis is performed as described in \citet{Serenelli:2013fz}.
In summary, if $\mathbf{v}$ is a set of stellar parameters from models 
(e.g. mass, radius, metallicity, age, effective temperature), and  
$\mathcal{O}$ the observed data, i.e.\,$\mathcal{O}
\equiv  \left(T_{\rm eff},  \Delta \nu,  \nu_{\rm max},  {\rm [Fe/H]}\right)$,
then according to Bayes' rule the probability density function (PDF) of
$\mathbf{v}$ given $\mathcal{O}$, i.e.\,the posterior PDF $p\left(\mathbf{v} |
\mathcal{O}\right)$, is given by
\begin{equation}\label{eqn:pdf} 
p\left(\mathbf{v} | \mathcal{O}\right) \propto p\left(\mathbf{v}\right)  
p\left( \mathcal{O}|\mathbf{v} \right), 
\end{equation} 
where $p\left(  \mathcal{O}|\mathbf{v}   \right)$ is the likelihood of 
$\mathcal{O}$ given $\mathbf{v}$, and $p\left(\mathbf{v}\right)$ is the prior 
PDF of $\mathbf{v}$ and represents prior knowledge on these quantities. The 
likelihood is computed assuming errors in $\mathcal{O}$ follow Gaussian  
distributions. 

For determining $p\left(\mathbf{v}\right)$ we considered the following. The 
effective selection function of pulsating stars observed by \emph{Kepler} is 
non trivial. Therefore, we cannot characterize a prior probability of stellar 
properties based on the fact that the star is in the \emph{Kepler} field. For 
this reason, we assume a flat prior in [Fe/H] and age including only a strict 
cut on the latter at 16~Gyr. For the prior in mass we assume a standard 
Salpeter IMF. In the cases where it is possible to measure the separation 
between mixed modes (see Section~\ref{sec:glob_astero} above), the information 
obtained on the evolutionary phase of the star is also included as a (binary) 
prior such that only stellar models corresponding to the determined 
evolutionary phase are used in the analysis. When this information is not 
available all models are considered and stellar properties can suffer from large
uncertainties due to both the RGB and clump evolutionary phases having 
non-negligible likelihoods. Bimodal probability distributions could in 
principle arise in those cases, influencing the estimation of reliable 
properties for these stars. 

Once the posterior PDF has been computed for a star, the posterior PDF of any
stellar quantity $x$ can be simply obtained as
\begin{equation}\label{eqn:poster} 
p\left(x|\mathcal{O}\right) = \int{\delta(x(\mathbf{v})-x) p\left(\mathbf{v} |
  \mathcal{O}\right) }w_v d^3v, 
\end{equation}
where $w_v$ accounts for the volume of parameter space occupied by an
isochrone point characterized by $\mathbf{v}$ \citep[see appendix A in][]{c11}.

The computation of the likelihood function in Equation \ref{eqn:pdf} requires 
determination of theoretical values for $\Delta\nu$ and $\nu_\mathrm{max}$. 
Those can be derived from the adopted grid of isochrones thanks to scaling 
relations. To a good approximation, $\Delta\nu$ scales with the mean density of 
the star \citep{Ulrich:1986ge}, while $\nu_\mathrm{max}$ is related to the
 surface properties \citep{brown91,Kjeldsen:1995tr}. From the solar values, two 
asteroseismic scaling relations can be written 
\citep[e.g.,][]{Stello:2009gh,Miglio:2009hz}:
\vspace{0.2cm}
\begin{equation}\label{eqn:dnu} 
\frac{\Delta\nu}{\Delta\nu_\odot} \simeq \frac{(M/M_\odot)^{0.5}(\teff/T_{\mathrm{eff},\odot})^{3}}{(L/L_\odot)^{0.75}}, 
\end{equation}
\begin{equation}\label{eqn:numax} 
\frac{\nu_\mathrm{max}}{\nu_{\mathrm{max}_\odot}} \simeq \frac{M/M_\odot(\teff/T_{\mathrm{eff},\odot})^{3.5}}{L/L_\odot}\,, 
\end{equation}
where $\Delta\nu_\odot=$135.1~$\mu$Hz and $\nu_{\mathrm{max}_\odot}=$3090~$\mu$Hz 
are the values observed in the Sun for the adopted method in this paper 
\citep{Huber:2011be}. Using parallaxes and interferometric measurements, radii 
determined using these scaling relations have been shown to be accurate to 
better than 5\% in dwarfs \citep[e.g.,][]{vsa12,hub12,White:2013bu}. 
Further tests on the reliability of the scaling relations are summarized in 
\cite{mi13}.

Having now all the information needed to compute both the prior and the 
likelihood, the PDF for the relevant parameters of each star can be computed, 
including masses and radii. Further, precise distances can also be determined 
by scaling these radii with the angular diameters obtained via the IRFM.
As summarized in Section~\ref{sec:sp}, stellar parameters have been determined 
via an iterative process coupling the photometric and asteroseismic 
parameters. In every iteration we used the median and 
68\% confidence levels of the PDF to determine the central value and 
(asymmetric) uncertainty of each quantity. We call these errors ``formal 
uncertainties'', and we explore the effect of using different stellar models 
and pipelines on the error budget later in Section \ref{sec:err}.

\subsection{Unresolved binary contamination}\label{sec:bin}

In principle, the presence of undetected binaries is important since, should 
this happen, the colors measured for a star would also carry the contribution 
from its companion, thus affecting all photometric 
quantities derived and subsequently used for the seismic analysis (Sections 
\ref{sec:teff} - \ref{sec:astero}). Fortunately our sample comprises mostly 
RGB stars, for which the fraction of similar luminosity binaries is extremely 
rare \citep[$\sim1$\%, e.g.,][]{nataf12}. 
Moving to lower mass companions, the fraction of spectroscopic binaries 
among K and M giants ranges from $\sim 6$ to $\sim30$\%, where observational 
difficulties as well as accounting for selection effects prevent better 
estimates \citep[e.g.,][]{f09}. 

A number of important photometric flags are available for all seismic 
stars used in this work, $96$\% of our targets having the best 
possible 2MASS pedigree$^9$: quality flag ``AAA'', blend flag 
``111'' (meaning one component is fit to the source) and contamination 
and confusion flag ``000'' (meaning the source is unaffected by known 
artifacts) in all three infrared 
bands. The remaining $4$\% of stars with one or more flags set differently from 
above, usually have larger photometric errors which thus are accounted for in 
our Monte-Carlo, naturally resulting in larger error bars (see Section 
\ref{sec:err}).

Synthetic photometry offers a qualitative way of assessing the level of 
contamination expected from a companion. 
From the MARCS library of synthetic spectra \citep{g08} we assume 
$\logg=2.5, \feh=0$ and $\teff=4750$~K for a typical giant to which we 
assign a radius of $10\,\rm{R}_{\odot}$. For the secondary, we consider two 
representative cases: in one the companion is identical to the Sun, 
while in the other it is an M dwarf ($\logg = 5.0, \feh=0$ and $\teff=4000$) 
with radius $0.7\,\rm{R}_{\odot}$. Assuming that both binary components randomly 
sample the initial mass function, this last case is likely to be the most 
probable, given that by number M dwarfs are the dominant stellar population in 
the Galaxy \citep[e.g.,][]{reid_book}.
 
As expected, the flux contribution of an M dwarf is entirely negligible 
compared to a giant: optical and infrared colors are affected by 1 to 4 
millimag, thus having no impact on the IRFM. Str\"omgren $b-y$, $m_1$ and 
$c_1$ indices are even less affected, a few parts over ten thousand at most. 
In the presence of a solar companion, the effect is between $0.01$ and $0.02$ 
magnitudes in both 
broad-band and Str\"omgren colors, with only moderate impact on the 
photometric parameters derived. Our conservative error bars 
account for any such uncertainty. Of course, other combinations in the ratio 
of the primary to secondary effective temperature and radius are 
possible, but the two cases explored above are representative of the most 
common scenarios, and give us confidence that binary contamination, when 
present, barely affects our parameters.

{\it Kepler} data provides an additional tool to quantify binary contamination 
in the sample. In addition to $\Delta\nu$ and $\nu_{\rm{max}}$ the 
asteroseismic analysis described in Section \ref{sec:glob_astero} yielded 
estimates of the global oscillation amplitude, which is well-known to correlate 
with stellar properties \citep[see, e.g.,][]{corsaro13}. Any near equal-mass 
binary companion would result in a second oscillation signal with nearly equal 
frequency, causing a significant dilution of the observed amplitude. A second 
oscillation signal is observed for only five stars in our sample, with 
estimated luminosity ratios ranging from $\simeq 0.4-0.8$. These stars have 
been flagged in the analysis. 

\subsection{Total error budget}\label{sec:err}

For all 9 possible different combinations of broad-band photometric systems 
(three) and reddening maps (three), we derive both self-consistent parameters 
and uncertainties.
For each star, uncertainties in the physical parameters derived from the IRFM 
(effective temperature, bolometric flux and angular diameter) are 
computed in a fashion very similar to \cite{c10}. First, a Monte-Carlo is run 
to account for random photometric errors, according to the set of filters used 
in the IRFM. We use uncertainties quoted for each band in 2MASS ($JHK_S$) and 
APASS ($BV$ and $g'r'i'$); no star-by-star uncertainties are reported 
in the KIC ($griz$), and we thus use a constant uncertainty of $0.02$~mag in 
each filter \citep[figure 4 in][]{kic}. 

\begin{figure}
\epsscale{1}
\plotone{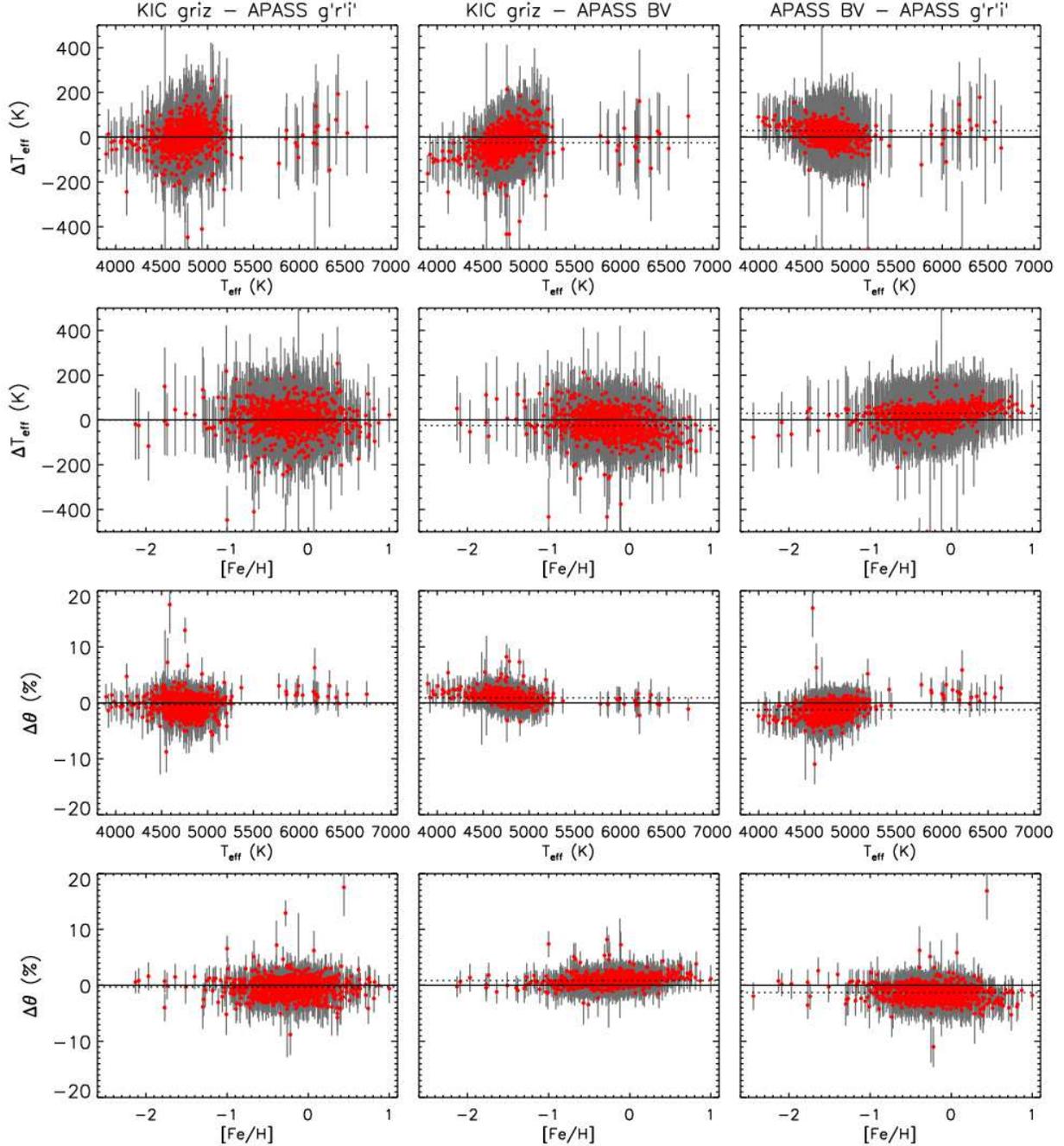}
\caption{Comparison between $\teff$ (first and second rows) and $\theta$ 
(third and fourth rows) derived implementing different optical photometric 
systems in the IRFM (first, second and third column), with corresponding error 
bars. 
In all instances the same reddening map is used (see discussion in Section 
\ref{sec:err}) although adopting a different one barely changes this 
comparison. Dotted lines represent the median differences. 
\label{f:upj}}
\end{figure}

For reddening we explore a systematic variation of $\ebv = \pm 0.02$ with 
respect to the values adopted from each map. 
This variation in reddening is also used to compute by how much the Str\"omgren 
metallicities change. As already mentioned (Section \ref{sec:feh}), this 
metallicity variation is added quadratically to the scatter of the $\feh$ 
calibration, returning a typical global uncertainty of $0.17$~dex. This is 
used to compute how much the parameters derived via the IRFM vary, although we 
recall once again that the IRFM is only mildly sensitive to the input $\feh$. 
In this way, reddening enters our errors twice, both in the broad-band 
colors and in the metallicities used in the IRFM; the choice to maximize its 
effect is motivated by the fact that we aim to derive realistic and 
conservative uncertainties. Finally the effect of varying $\logg$ by $0.1$~dex 
is included: the precision of seismic gravities is actually much higher 
\citep[$\lesssim 0.03$~dex, e.g.,][]{gai11,thy12,morel12} but the IRFM is so 
largely insensitive 
to $\logg$ that even this very conservative uncertainty has virtually no 
effect on the error budget. All uncertainties listed above (Monte-Carlo, 
reddening, metallicity and gravity) are added quadratically, and 
further increased by $1.0$\% and $0.7$\% for bolometric fluxes and angular 
diameters and $20$~K for $\teff$ to account for our uncertainty in the absolute 
zero-point of those scales \citep{c10}. 
We recall that the effective temperature, bolometric flux and angular diameter 
of each star are usually derived using three different implementations of 
the IRFM ($grizJHK_S$, $BVJHK_S$ and $g'r'i'JHK_S$): comparison among those is 
shown in Figure \ref{f:upj}. For the sake of the latter, the same reddening 
map is used although in some cases distances change depending on the filter 
combination used in the IRFM, and thus the values derived for $\ebv$ change as 
well, even when using the same map.
Metallicities are derived from Str\"omgren colors only, although for the 
reason just explained $\ebv$ might be slightly different among the three 
implementations of the IRFM. Fortunately this reddening variation {\it within} 
the same map has virtually no effect on $\feh$, often being null and typically 
well below $0.01$~dex.

Having now three different sets of $\fbol$, $\theta$, 
$\teff$ and $\feh$ each with its own error, we can combine this information by 
using the weighted average. The values derived in this way for a given 
reddening map are dubbed ``consolidated'' parameters.
In the ideal case, different filter 
combinations should all return the same values for a given reddening map. 
Within the uncertainties, this is indeed the case (Figure \ref{f:upj}). The 
weighted 
sample standard deviation, which measures the overdispersion of the weighted 
averaged parameters, tells us their internal consistency: reassuringly, 
this is very close to the values obtained with the 
Monte-Carlo for each implementation of the IRFM. 
To obtain the final global errors, we quadratically add the weighted sample 
standard deviation to the errors computed above for each star.
Their median values are $82$~K in $\teff$, $0.17$~dex in $\feh$, $2$\% in 
angular diameters and $5$\% in bolometric fluxes. 

\begin{figure}
\epsscale{1}
\plotone{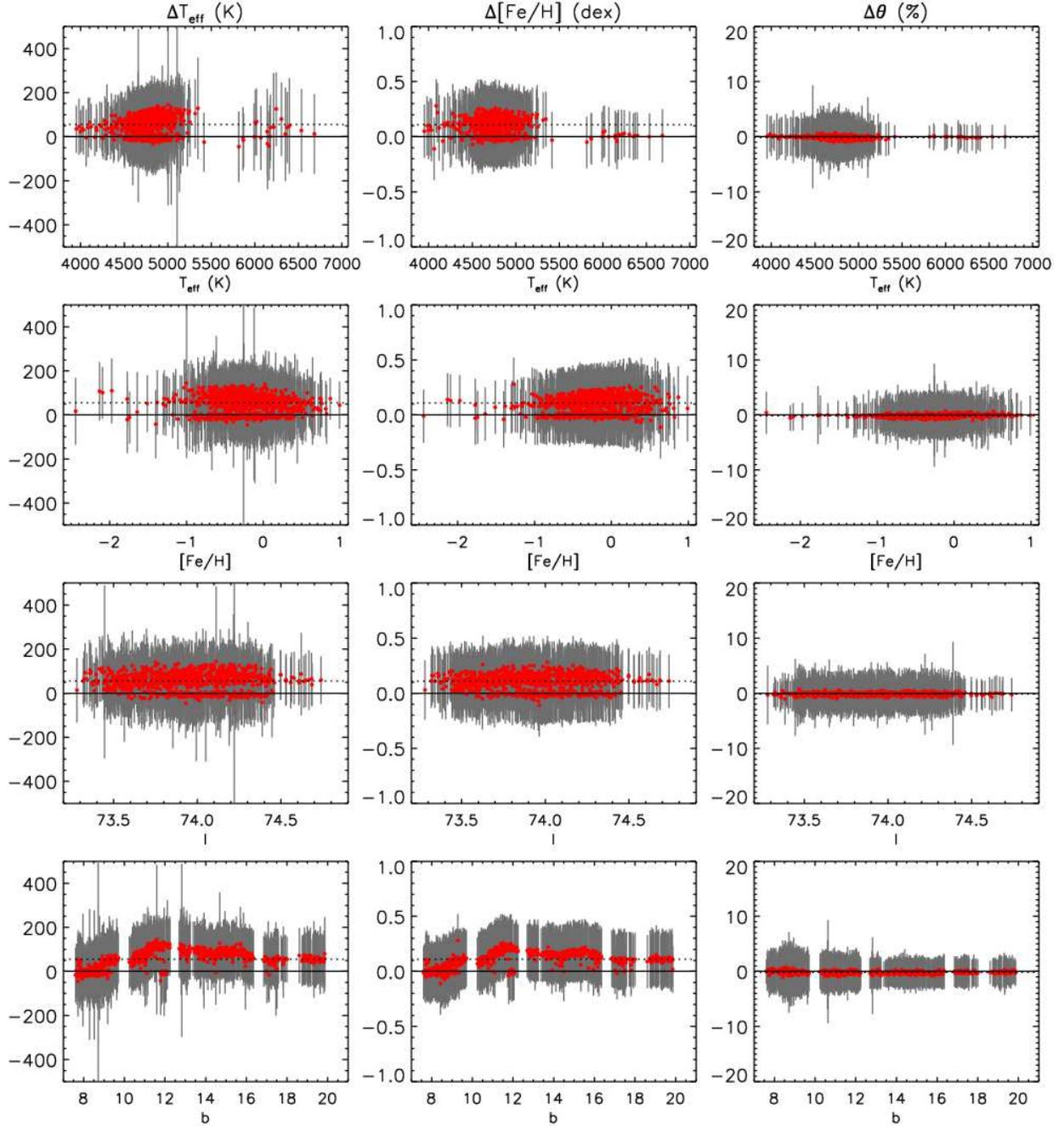}
\caption{Comparison between the consolidated set of effective temperatures 
(first column), metallicities (second column) and angular diameters (third 
column) derived 
using two different reddening maps (\citealt{al05} minus \citealt{drimmel03}). 
Stars showing almost no difference are those located at Galactic latitudes 
$b\simeq8.5$, where both maps are anchored at the $\ebv$ of NGC\,6819. Dotted 
line is the median difference.
\label{f:da}}
\end{figure}

The procedure outlined above is repeated for each reddening map. Comparison 
between the results obtained adopting \cite{drimmel03} or \cite{al05} is 
shown in Figure \ref{f:da}: as expected from our conservative approach in 
estimating realistic error bars, results always agree within the uncertainties. 
Median differences are relatively small and rather uniform with no 
particular dependence on Galactic coordinates, especially for latitudes above 
$10^{\circ}$. At smaller $b$ the agreement is much better and reflects our use 
of the reddening of NGC\,6819 as an anchor point. 

\begin{figure}
\epsscale{0.93}
\plotone{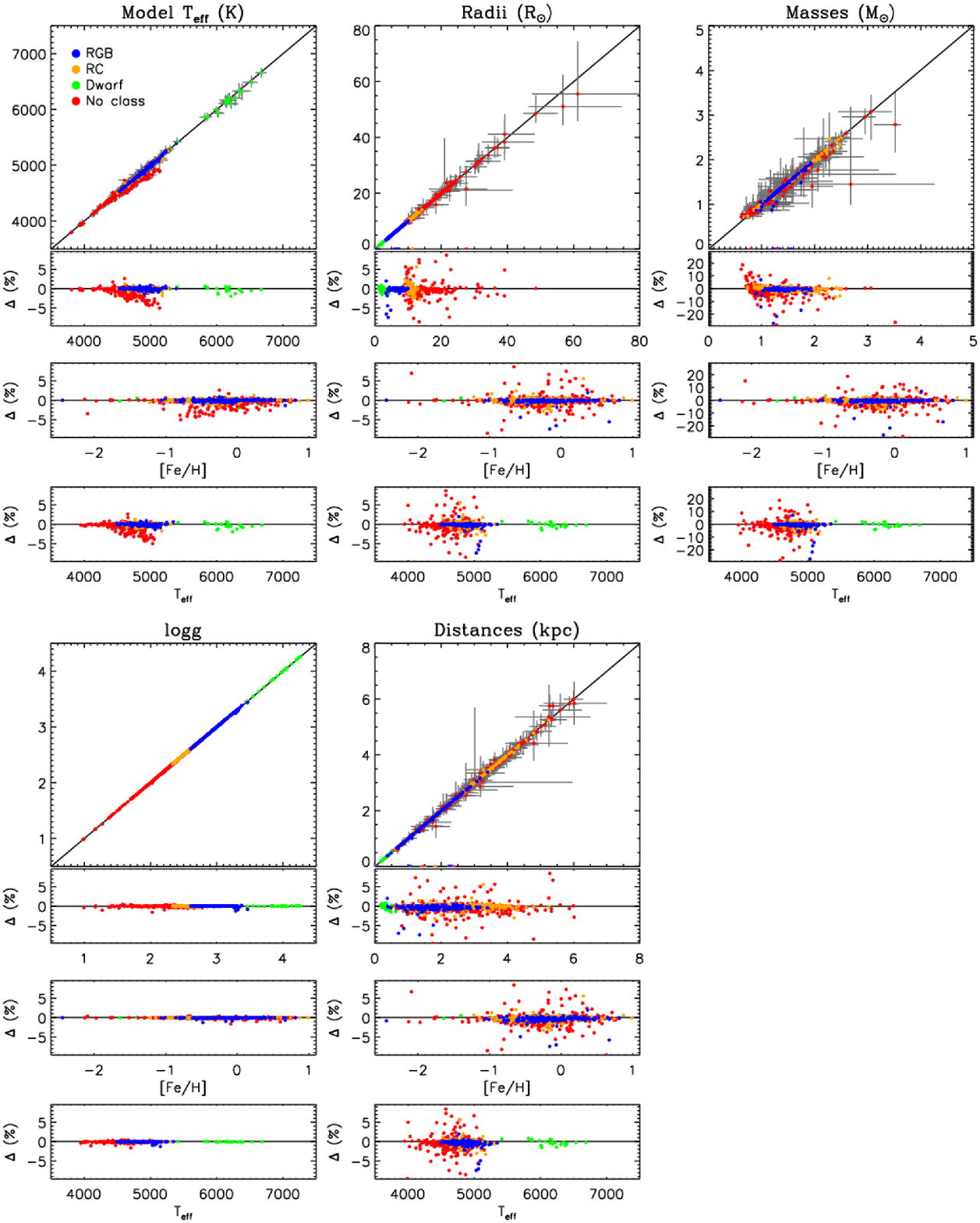}
\caption{Difference in the classical plus seismic quantities derived when 
using the Am\^ores \& L\'epine (2005, horizontal axis) or Drimmel et.~al (2005, 
vertical axis) reddening maps. 
The fractional difference is shown as function of the parameter 
investigated, as well of the consolidated $\teff$ and $\feh$. Different colors 
identify stars with different seismic classification.
\label{f:ad}}
\end{figure}

These consolidated parameters are used once again into the seismic scheme 
described in Section \ref{sec:astero} to derive a final set of radii, 
distances and masses with median formal uncertainties of order $2$\%, $3$\%, 
and $5$\% respectively. These values are of the same order as those found 
in other asteroseismic studies when reliable information on stellar effective 
temperatures and metallicities are available. In particular, errors in 
masses and radii are positively correlated, and this leads to very small 
errors in gravities \citep[e.g.][]{gai11,creev13,chap13}. Nevertheless, it 
should be kept in mind that as of yet, there are only few empirical tests of 
the accuracy of the scaling relations when applied to giants (see also 
discussion in Section \ref{sec:astero}), and modest systematic errors are thus 
possible \citep{mbs12}. However, the clear separation of the first and 
secondary clump in our data gives confidence on the internal precision of our 
gravities. 
Figure \ref{f:ad} extends the comparison between different reddening maps to 
seismic quantities, where stars with different seismic classification are 
highlighted in colors; differences are always within the error bars also in 
this case. 
We also explore the impact in the derived stellar parameters of varying the 
input physics in our reference BaSTI grid of isochrones. We test the 
effect of switching on mass-loss with high efficiency ($\eta=0.4$) during the 
evolution, finding that the formal uncertainties 
quoted above are more than two times larger than the median difference in the 
derived stellar parameters. Similarly, isochrones with no core overshoot 
during the main-sequence phase give differences about ten times smaller than 
the formal uncertainties. Hence, these effects have overall little impact on 
the parameters derived here (although this would not always be the case for 
stellar ages).

\begin{figure}
\epsscale{0.93}
\plotone{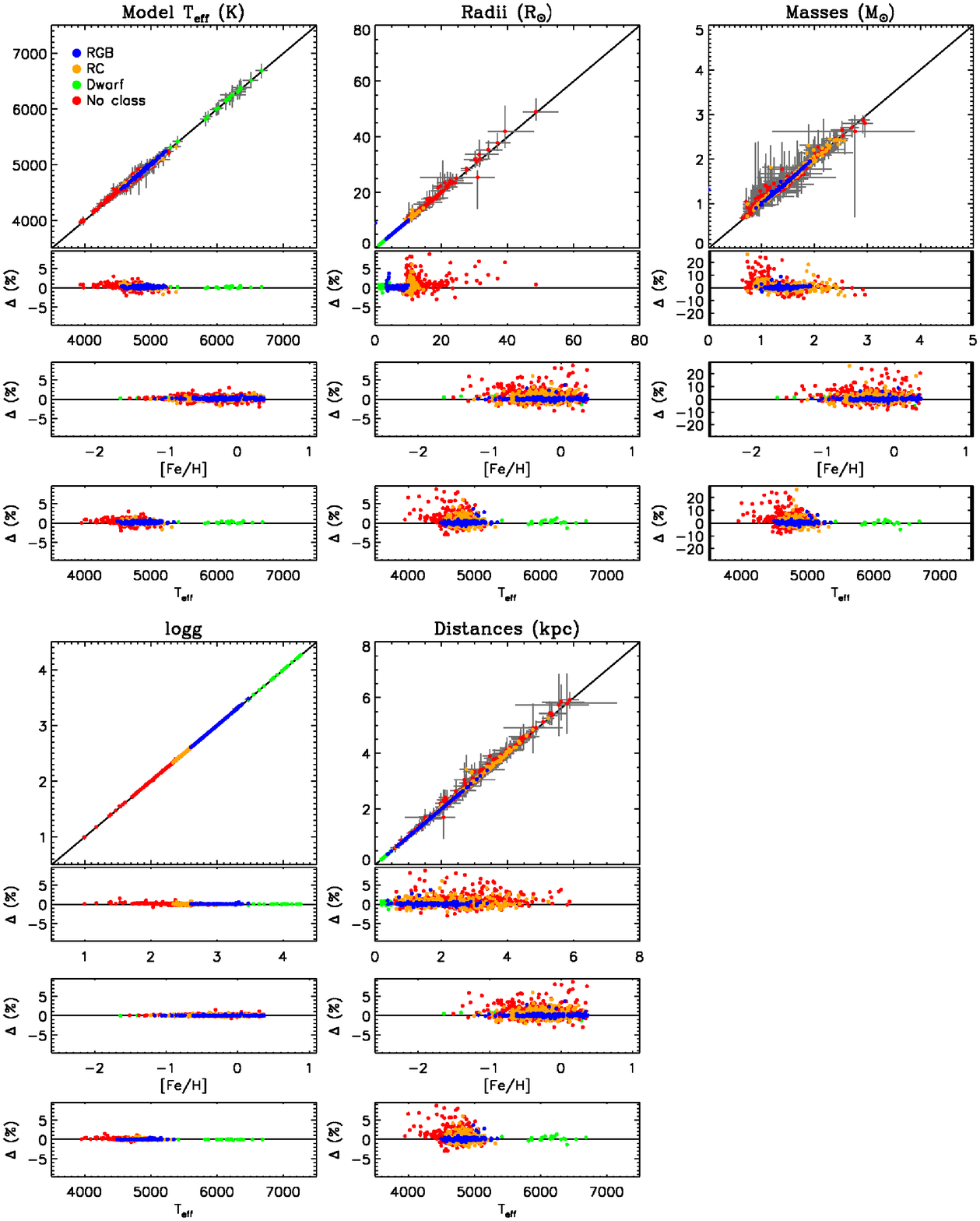}
\caption{Same as Figure \ref{f:ad}, but comparing the effect of using a 
Bayesian (horizontal) or a Monte-Carlo (vertical) scheme with the BaSTI 
models to derive seismic parameters. The same reddening map is assumed in 
both cases.
\label{f:huber}}
\end{figure}

To account for systematics arising from different evolutionary codes and 
implementation of input physics we derived Bayesian stellar parameters using 
and extension of the grid of models presented in \citet{Serenelli:2013fz}, 
constructed using the GARching STellar Evolution Code \citep[GARSTEC,][]{ws08} 
and covering a mass and metallicity range of $0.6\le$M$_\odot\le3.0$ and 
$-3.0\le\feh\le+0.5$. Similarly, we 
determined stellar properties from BaSTI evolutionary tracks using the 
Monte-Carlo approach described in \cite{huber13}, to explore systematics 
introduced by our Bayesian pipeline. Comparison between our reference 
parameters and those derived with a different approach are shown in 
Figure \ref{f:huber}. For stars with known seismic 
classification differences are usually only a few percent, while they can grow 
considerably larger for stars without seismic classification. 

\begin{figure}
\epsscale{1}
\plotone{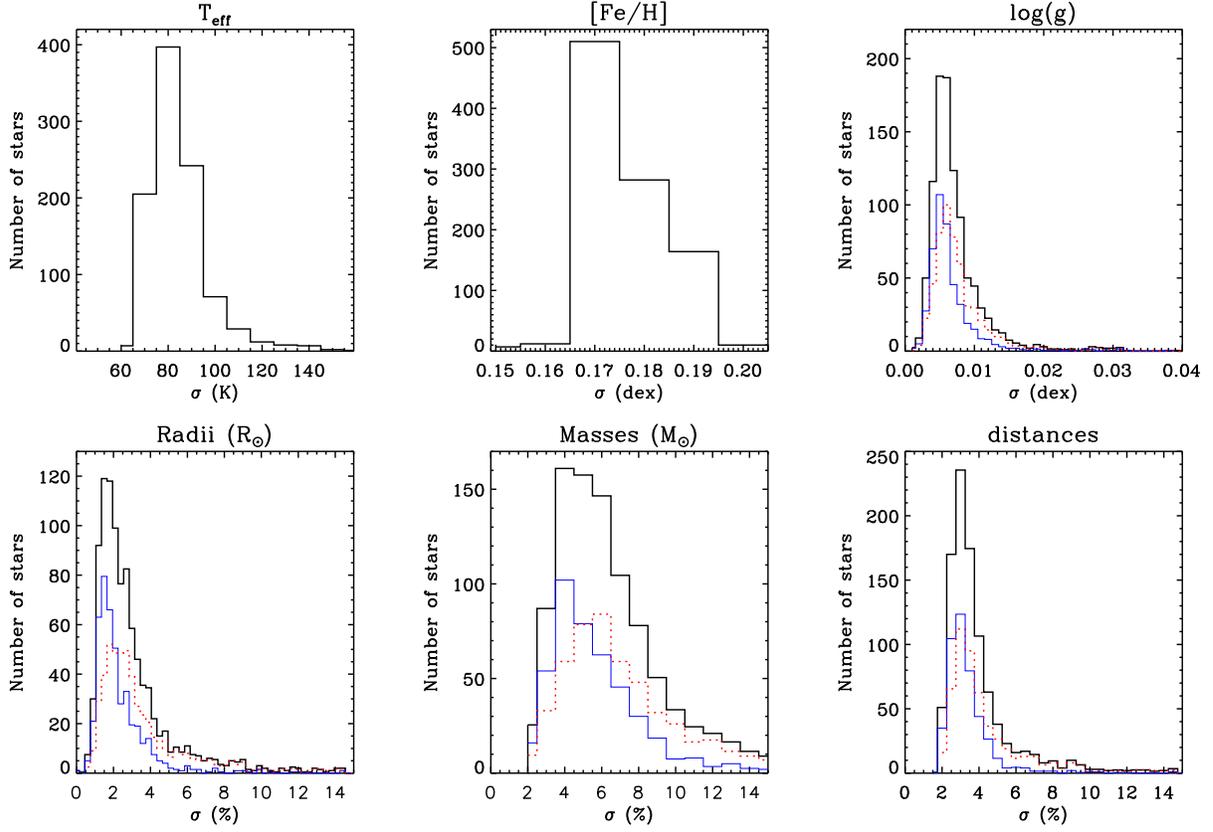}
\caption{Final error distributions of the main stellar parameters derived. 
For the asteroseismic quantities, the overall distribution is also split into 
stars with (blue) and without (red) evolutionary phase classification (Section 
\ref{sec:glob_astero}).\label{f:her}}
\end{figure}

For the final and global uncertainties in $\logg$ (and densities), radii, 
distances and masses, we add quadratically to the formal uncertainties from 
our BaSTI reference models half the difference between those results and the 
ones obtained with the GARSTEC grid and the Monte-Carlo approach. 
Their median values are $82$~K in $\teff$, 
0.17~dex in $\feh$, $0.006$~dex in $\logg$, $1.5$\% in stellar density, 
$2.4$\% in radius, $3.3$\% in distance and $6.0$\% in mass. As expected, the 
uncertainty in asteroseismic quantities is also smaller for stars with known 
evolutionary phase classification compared to stars for which this 
information in unknown (Figure \ref{f:her}). In all instances, it should 
be kept in mind that while we have carefully accounted for various sources of 
uncertainties, the accuracy of the seismic parameters is currently limited by 
empirical tests of the scaling relations, which here are assumed to be exact.

\subsection{Adopted parameters}\label{sec:adopted}

As already described, stellar parameters have been derived self-consistently 
for 9 different combinations of photometric systems and reddening maps. 
We also verified in the previous Section that the use of three 
different photometric systems in the IRFM did not introduce any major 
difference in the results and hence merged those, thus curbing our 
combinations to the three reddening maps. For our field, the symmetric and 
spiral model of \cite{al05} yield virtually identical results (Section 
\ref{sec:red}), effectively meaning that we are dealing with 2 different 
sets of results. 

\begin{figure}
\epsscale{1}
\plotone{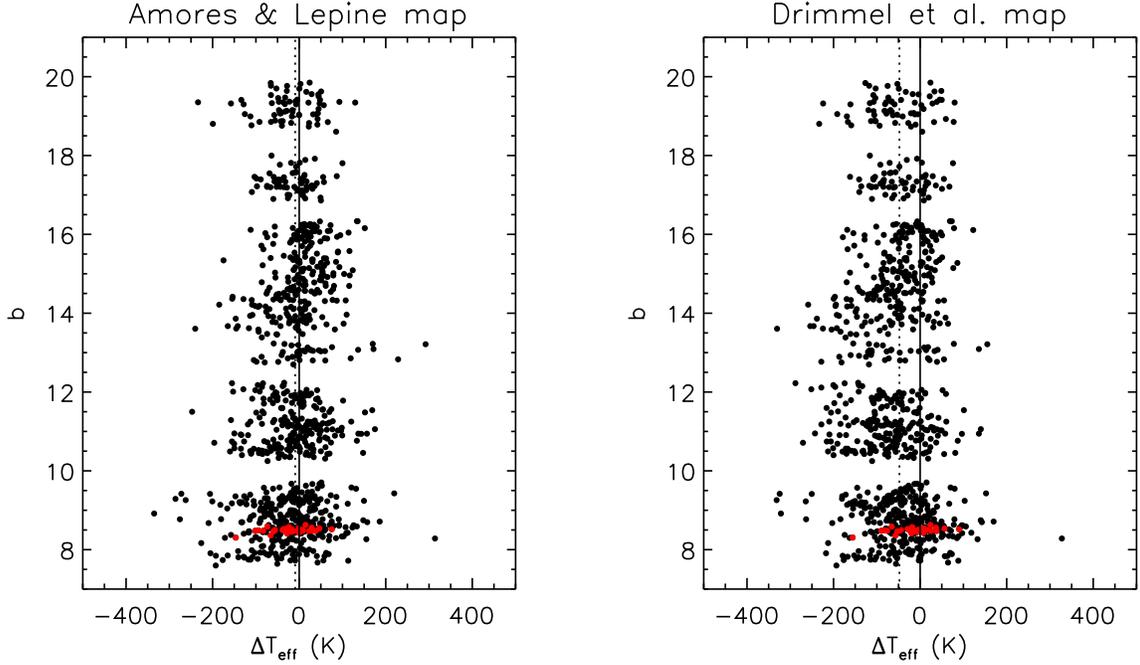}
\caption{Difference between $\teff$ obtained from the IRFM and from theoretical 
modelling for each star (IRFM minus models) as function of Galactic latitude 
$b$ for two different reddening maps, as indicated. Dotted line is the 
median difference. At low $b$, stars belonging to NGC\,6819 and used to 
calibrate both reddening maps are highlighted (see Section \ref{sec:red}).
\label{f:modelred}}
\end{figure}

\begin{figure}
\epsscale{1}
\plotone{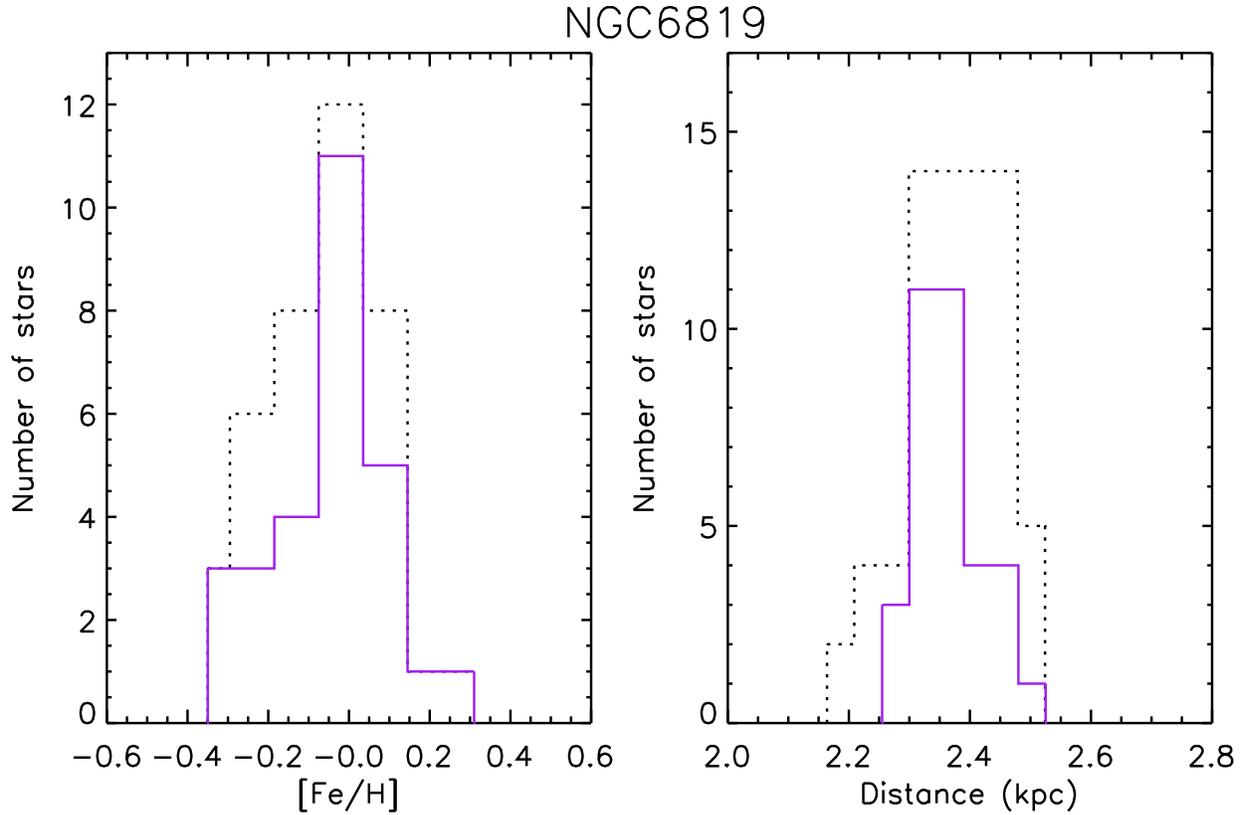}
\caption{Parameters determined for giants in the open cluster NGC\,6819, 
selected 
according to the seismic classification of \cite{stello11}. {\it Left panel:} 
metallicities for all stars (dotted line) or retaining only stars with good 
photometric quality flag (continuous line, Section \ref{sec:sspc}). 
{\it Right panel:} distances for the same stars. Now, dotted line refers 
to all stars, while continuous line is for stars having certain seismic 
classification (Section \ref{sec:glob_astero}).
\label{f:6819}}
\end{figure}

\begin{figure}
\epsscale{0.76}
\plotone{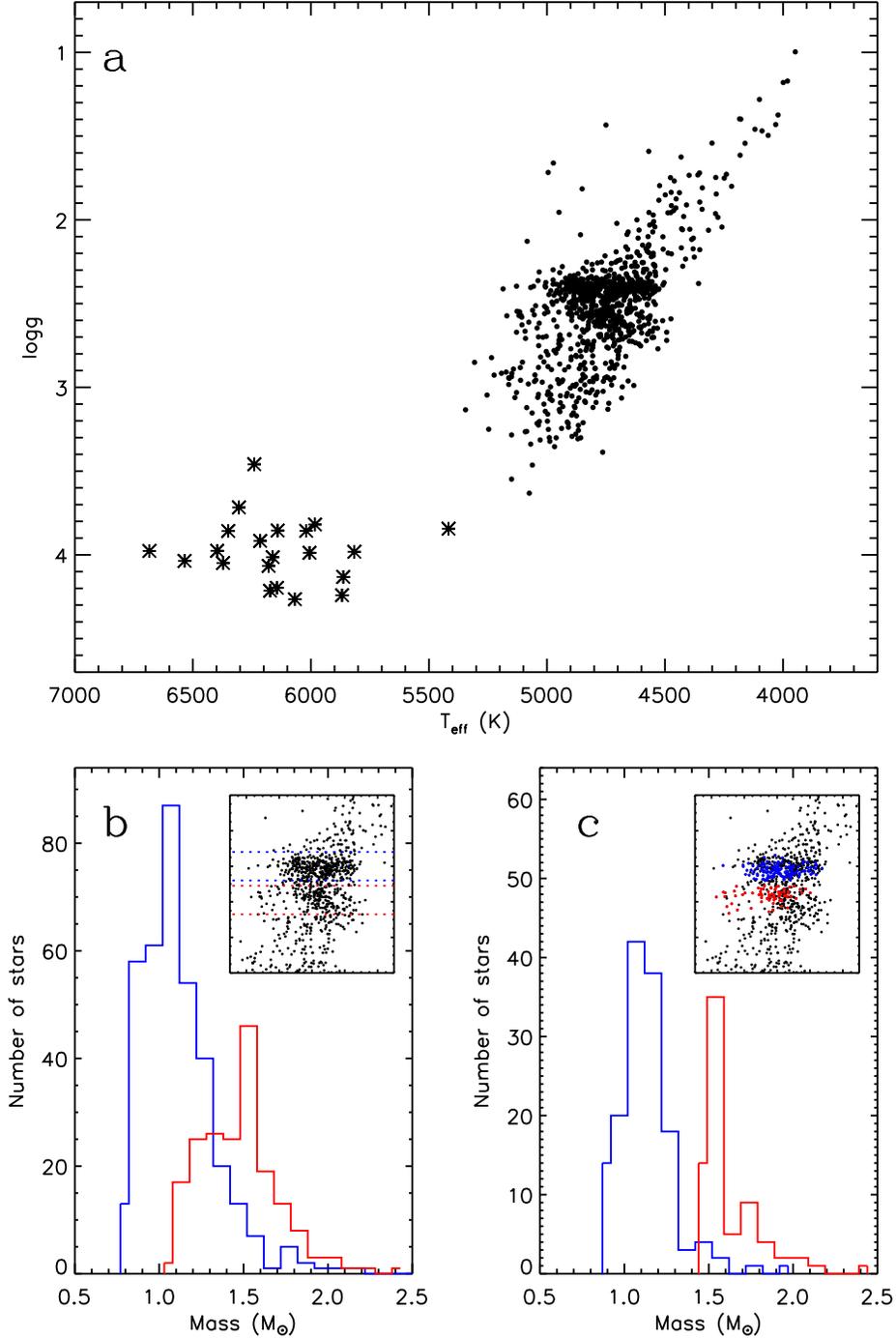}
\caption{{\it Panel a:} $\teff$ and $\logg$ for our sample of stars. 
Asterisks/circles identify stars for which we used the dwarfs/giants 
metallicity calibrations (Section \ref{sec:feh}).
{\it Panel b:} first (blue) and secondary (red) clump dissection based only on 
$\logg$ selection (range indicated by dotted lines in the upper 
inset) and corresponding histogram for their mass distribution. 
{\it Panel c:} same as previous panel, but retaining only stars with 
certain clump classification.
\label{f:fs}}
\end{figure}

Figure \ref{f:modelred} compares the model ($\mathbf{v}$ in 
Equation \ref{eqn:pdf}) and empirical effective temperatures (from the IRFM) 
derived with the two reddening maps. Notice that the effective temperature is 
only one of many parameters entering our Bayesian scheme, and in fact 
the model effective temperature barely changes when adopting a different 
reddening map
(see top left panel in Figure \ref{f:ad}, where the mean and median difference 
is only 12 and 2~K, respectively). While differences might exist between 
the empirical and the model $\teff$ scale, the $\Delta\teff$ trend with 
respect to Galactic latitude in the \cite{drimmel03} case (Figure 
\ref{f:modelred}, right panel) reminds that of Figure \ref{f:red} (lower 
right panel), thus dictating our preference for \cite{al05}. Notice though, 
our reddening errors account for the difference between the two maps (Section 
\ref{sec:err}).

To summarize, our photometric stellar parameters are derived adopting 
the \cite{al05} reddening map (rescaled to match the reddening of NGC\,6819) 
with seismic parameters obtained using the BaSTI non-canonical isochrones with 
no mass-loss and the global errors derived in Section \ref{sec:err} and shown 
in Figure \ref{f:her}. 
The catalogue is available in the electronic version of the journal, and Table 
\ref{t:t1} provides information about each tabular column.
Depending on the purpose for which we use our parameters, we can 
also restrict our analysis to include only stars with more certain seismic 
classification.

\subsection{External validation of derived parameters}

Throughout the paper we have used the open cluster NGC\,6819 as a benchmark 
point. Here we used it to further check the consistency of the main 
parameters derived from the seismic analysis. Figure \ref{f:6819} shows the 
metallicity distribution of the seismic giants in the cluster: just as 
expected from the discussion of Section \ref{sec:feh} the solar metallicity 
of this cluster is reproduced. More interesting is the second panel, 
which shows the distance distribution for the same stars. Their distance 
sharply peaks at a value of $2.38 \pm 0.08$~kpc, in excellent agreement with 
similar estimates obtained from isochrone fitting \citep[e.g.,][]
{yang13,sand13}. 

Figure \ref{f:fs} shows all seismic targets analyzed in this work in the 
empirical (IRFM) $\teff$ -- seismic $\logg$ plane: the very clear presence of 
primary and secondary clump stars appear. The latter population was discovered 
by \cite{girardi98} using {\it Hipparcos} parallaxes and its detection here 
clearly illustrates the precision we have achieved. Such a population is 
composed by stars with $M\gtrsim 1.8~M_{\odot}$ that ignite helium in 
non-degenerate conditions and it is relatively short-lived, peaking at 
$\sim1$~Gyr \citep[e.g.,][where the exact mass and age depend on the 
adopted models]{girardi99}.

\section{Comparison with other works}

\subsection{KIC}\label{sec:kic}

Currently, the largest database of stellar parameters for stars in the {\it 
Kepler} field is provided by the Kepler Input Catalogue \citep{kic} 
which was originally conceived as a tool to optimize the target 
selection \citep{bata}, and thus provide only approximate estimates of stellar 
parameters. A number of investigations has highlighted inaccuracies in 
its parameters \cite[e.g.,][]{mz11,pi12,thy12}.

\begin{figure}
\epsscale{0.99}
\plotone{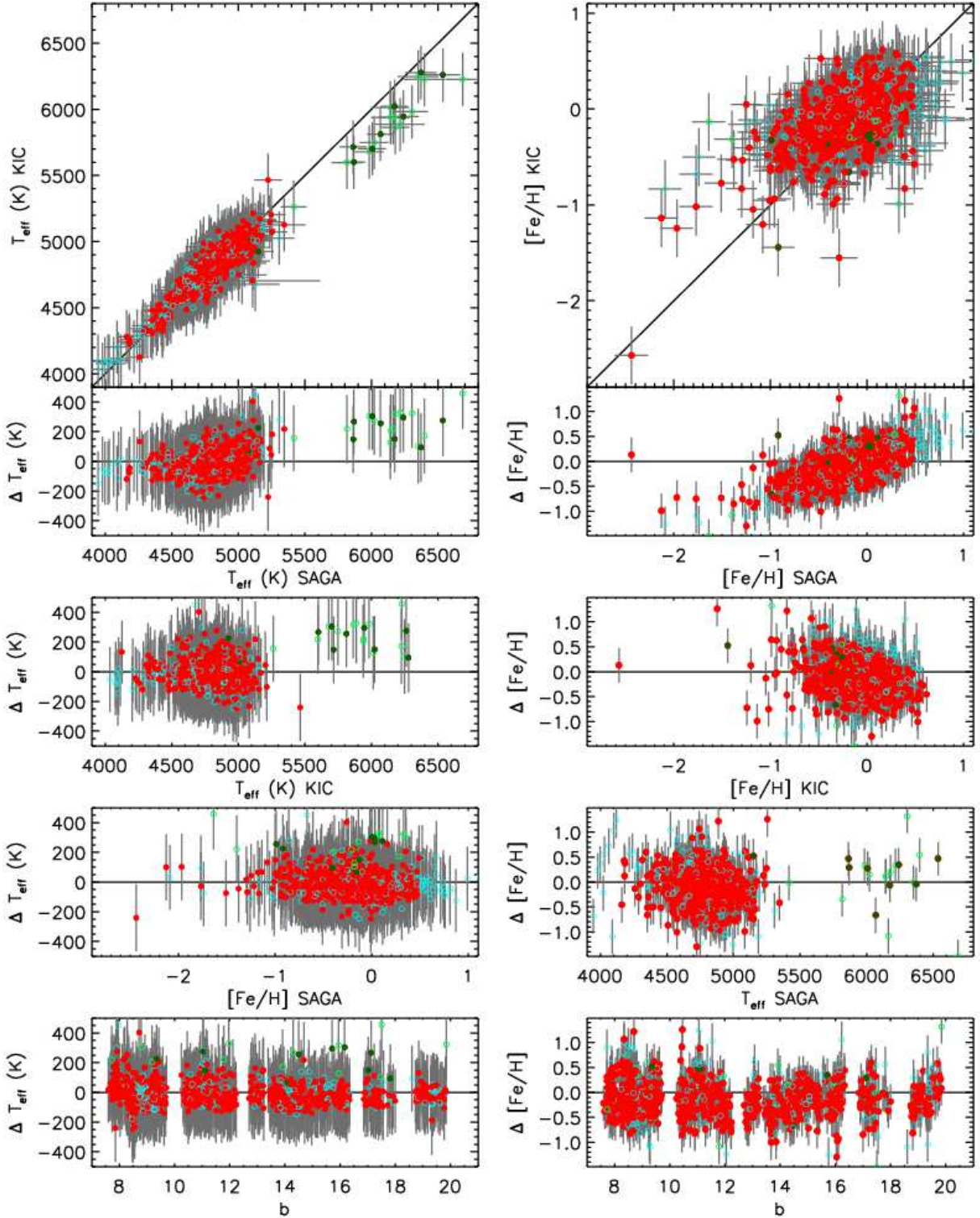}
\caption{Comparison between $\teff$ (first) and $\feh$ (second column) 
determined for stars in the SAGA and KIC ($\Delta=$ SAGA minus KIC). 
Filled-red/open-cyan (filled-dark/open 
green) circles identify giants (dwarfs) with more certain/uncertain 
Str\"omgren labels (see Section \ref{sec:feh}). 
\label{f:kic}}
\end{figure}

In Figure \ref{f:kic} we compare the SAGA $\teff$ and $\feh$ against the KIC. 
The effective temperatures of giants are in overall reasonable agreement (KIC 
being some $20$~K hotter but with a large $80$~K scatter), also considering 
uncertainties related to reddening. However, a clear offset appears when 
moving to main-sequence stars, spurring systematic differences by more than 
$200$~K (SAGA being hotter). These trends are in agreement with similar 
findings for dwarfs by \cite{pi12} and for giants by \cite{thy12}.

Concerning $\feh$, the KIC performs less well. For giants 
$\Delta \feh = -0.13$~dex (SAGA minus KIC) and $\sigma = 0.35$~dex, which 
improves only moderately if restricting to stars with the most reliable 
Str\"omgren metallicities. Part of the scatter might arise from the fact that 
metallicities in the KIC are more representative of the overall metal content 
[M/H], while we calibrated our photometry using direct $\feh$ 
measurements. Field stars in the Galactic disk are known to generally increase 
their $[\alpha/\rm{Fe}]$ content going towards lower metallicities 
\citep[e.g.,][]{tinsley79,mcwilliam97,matteucci01,pagel09}. If we assign 
$[\alpha/\rm{Fe}]$ to our stars according to the analytic model of 
\cite{pt95} and compute their [M/H], the offset and the scatter reduce to 
$-0.07$~dex and $0.31$~dex, respectively, 
with a particular improvement for stars with metallicities below $-1$ dex. 
A similar trend towards low metallicities in the KIC can also be seen in the 
comparison of \cite{dong13}.

\begin{figure}
\epsscale{1}
\plotone{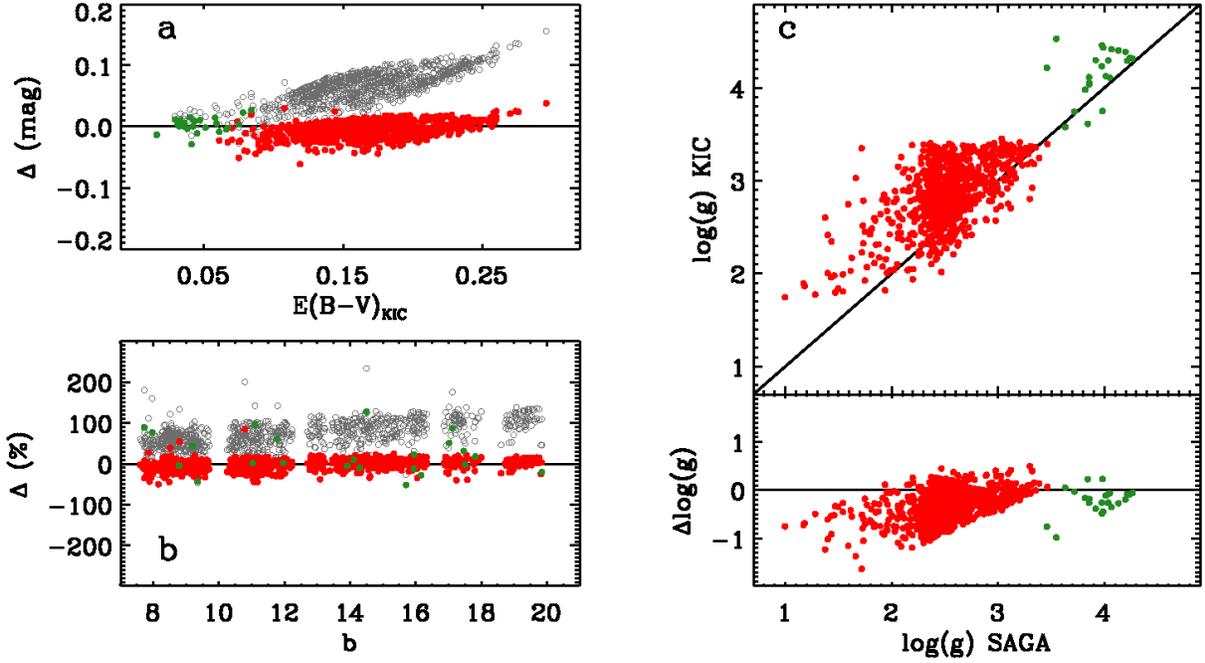}
\caption{{\it Panel a:} comparison between the color excess $\ebv$ in the KIC 
and SAGA (gray circles, KIC minus SAGA). Colored circles (green for dwarfs, 
red for giants) highlight the same comparison after correcting the KIC 
reddening according to Equation \ref{eq:ebvkic}. {\it Panel b:} same as in 
panel a, but showing the percent difference (KIC/SAGA) as function of Galactic 
latitude.
{\it Panel c:} comparison between $\logg$ in the KIC and asteroseismic values 
in SAGA.
\label{f:ebvlog}}
\end{figure}

It is also interesting to compare our results with the reddening and 
surface gravities reported in the KIC. Confidence in the SAGA values of $\ebv$ 
comes from the fact that the good agreement of our photometric $\teff$ with 
respect to spectroscopy (next Section) disappears if the reddening of the KIC 
is adopted. Our calibrated reddening map indicates that the color excess 
reported in the KIC is increasingly overestimated for increasing values of 
$\ebv$. 
The correlation shown in Figure \ref{f:ebvlog}a 
also appears when the same quantity on the $y$-axis is plotted as function of 
apparent magnitudes or distances (which in turn are highly correlated); thus 
indicating that fainter and further away objects in the KIC are more likely to 
have inaccurate values of reddening. This suggests that the scale height 
and/or the attenuation adopted in the KIC dust model are the primary causes for 
this disagreement. 
Another spatial dependence appears when plotting the percent difference as 
function of Galactic latitude $b$ (Figure \ref{f:ebvlog}b). At higher 
$b$ reddening decreases, but percentage-wise the KIC overestimation is 
larger. A linear fit of the trend in Figure \ref{f:ebvlog}a accounts 
for most of the difference between the KIC and SAGA, but it does not remove the 
trend of the fractional error as function of $b$. This is accounted for by 
fitting the ratio between the color excess in SAGA and the values obtained 
from the linear fit above. The corrected reddening we derive for the color 
excess in the KIC is thus:
\begin{equation}\label{eq:ebvkic}
\ebv = [0.55\,E(B-V)_{\rm{KIC}} + 0.008] (1.2 - 0.02\,b),
\end{equation}
where the first term is dominant, while the second accounts for the spatial 
dependency of the fractional difference. Since all reddening values in the KIC 
are derived using the same dust model, the correction proposed here likely 
applies for most of the stars in the {\it Kepler} field. It should be kept in 
mind that Equation \ref{eq:ebvkic} is tailored to the stripe analyzed in this 
work.

Figure \ref{f:ebvlog}c compares the surface gravities we derived from 
asteroseismology with the values reported in the KIC. The sharp cut at 
$\logg \sim 3.4$ reflects the KIC-based pre-selection of giants for 
measuring oscillations \citep[][]{ciardi11,hekker11b}. We 
have already discussed in Section \ref{sec:err} the precision of 
asteroseismic gravities; $\logg$ values in the KIC were derived with the 
purpose of broadly distinguishing between dwarfs and giant, and we confirm 
here the overall success in doing this \cite[but see][for a more 
detailed discussion]{h14}.

\subsection{APOGEE}\label{sec:apogee}

The APO Galactic Evolution Experiment (APOGEE) uses high-resolution 
$H$-band spectra to observe some $10^5$ giants across the Galaxy \citep{ahn13}. 
As part of this project, a number of targets in the {\it Kepler} field has 
been observed and are now publicly available via DR10. Here we compare our 
SAGA parameters with the spectroscopic ones from APOGEE. Details on the 
derivation of those can be found in \cite{meszaros13}. Briefly, two sets of 
parameters are provided: raw and corrected $\teff$ and $\feh$. Corrections for 
the latter are determined from fitting polynomials to reproduce a number of 
benchmark stars and clusters, and amount up to $0.2$~dex in metallicity 
and $100$~K in $\teff$ for the stars used here. The comparison carried out 
here is with respect to the corrected APOGEE parameters. 

The agreement in $\teff$ is good overall with $\Delta\teff = 90 \pm105$~K, SAGA 
being hotter. It is particularly good below $\simeq 4600$~K 
($\Delta\teff = 0 \pm34$~K) although APOGEE tends to increasingly 
underestimate $\teff$ descending the RGB towards hotter stars. 
Concerning metallicities, the comparison looks tighter because of the much 
higher quality of the APOGEE parameters with respect to the KIC. The same trend 
already highlighted in Figure \ref{f:kic} appears, although it is not as 
dramatic when we restrict the sample to stars with the best Str\"omgren 
photometry, $\Delta \feh = -0.11 \pm 0.26$~dex (only slightly larger than 
expected considering uncertainties in the SAGA and APOGEE values). While the 
raw APOGEE metallicities are more representative of the global [M/H], their 
corrected ones are forced to reproduce $\feh$ of a number of benchmark clusters 
\citep[see discussion in][]{meszaros13}. A few of these clusters are also used 
for our metallicity calibration: using for them the same $\feh$ adopted by 
APOGEE changes our metallicities at the level of $0.1$~dex toward the 
metal-poor regime.

\section{Conclusions}

In this paper we have presented Str\"omgren photometry of a stripe in the 
{\it Kepler} field. While the ultimate purpose is to use these observations to 
provide homogeneous and reliable stellar parameters for both candidate planet 
host and seismic stars, the geometry chosen for our observations already 
enables Galactic studies using stars with detected oscillations.

The strength of our approach is in the coupling of classical and 
asteroseismic parameters: effective temperatures (from the InfraRed 
Flux Method), metallicities (from Str\"omgren indices), masses, and radii 
(from seismology) are derived iteratively and self-consistently, thus giving 
access to other quantities such as reddening and distances, and providing
a complete picture of the seismic population in our observed Galactic fields.

The open cluster NGC\,6819 is located at the base of the observed stripe and 
offers an important benchmark to verify the soundness of our 
results. In particular, it allows us to anchor the metallicity scale of giants 
on the same zero-point used for investigating properties in the solar 
neighborhood via dwarfs and subgiants from the Geneva-Copenhagen Survey 
\citep{c11}. This is crucial if for example we wish to use nearby stars 
and/or giants (up to $\simeq 6$~kpc across the Galactic disk in this work) for 
the purpose of understanding the formation and evolution of stellar 
populations in the Milky Way.

The sample presented here covers latitudes between $8$ and $20$ degrees above 
the Galactic plane. The stellar parameters derived here are already well suited 
for a number of 
investigations; in particular the seismic classification allows us to 
distinguish between stars ascending the red giants branch burning hydrogen 
in a shell, and those which have also ignited helium burning in their core. 
Concerning the latter, the precision achieved allows us to discern between 
primary and secondary clump, hence whether the core ignites degenerately or 
not. 
With this information, and the parameters so far derived, we are thus in the 
position of using these stars to compute reliable stellar ages and 
investigate, for example the age-metallicity relation and the 
age/metallicity gradients across this part of the Galactic disk. Further, 
calibrating photometric metallicities and age dating techniques on the present 
sample, a deep all-sky Str\"omgren survey promises a leading role for Galactic 
studies.

\acknowledgments
We thank the referee for a prompt and constructive report.
This publication makes use of data products from the Two Micron All Sky Survey, 
which is a joint project of the University of Massachusetts and the Infrared 
Processing and Analysis Center/California Institute of Technology, funded by 
the National Aeronautics and Space Administration and the National Science 
Foundation. This paper makes use of data from the AAVSO Photometric All Sky 
Survey, whose funding has been provided by the Robert Martin Ayers Sciences 
Fund. Funding for the Stellar Astrophysics Centre is provided by The Danish 
National Research Foundation (Grant agreement No. DNRF106). The research is 
supported by the ASTERISK project (ASTERoseismic Investigations with SONG and 
{\it Kepler}) funded by the European Research Council (Grant agreement No. 
267864). D.H. is supported by an appointment to the NASA Postdoctoral Program 
at Ames Research Center, administered by Oak Ridge Associated Universities 
through a contract with NASA. A.M.S. is partially supported by the MICINN 
grant AYA2011-24704. This work has been supported by an Australian Research 
Council Laureate Fellowship to MA (grant FL110100012).

\appendix

\bibliographystyle{aj}
\bibliography{refs}

\clearpage

\clearpage

\clearpage

\clearpage

\clearpage

\begin{figure}
\epsscale{0.99}
\plotone{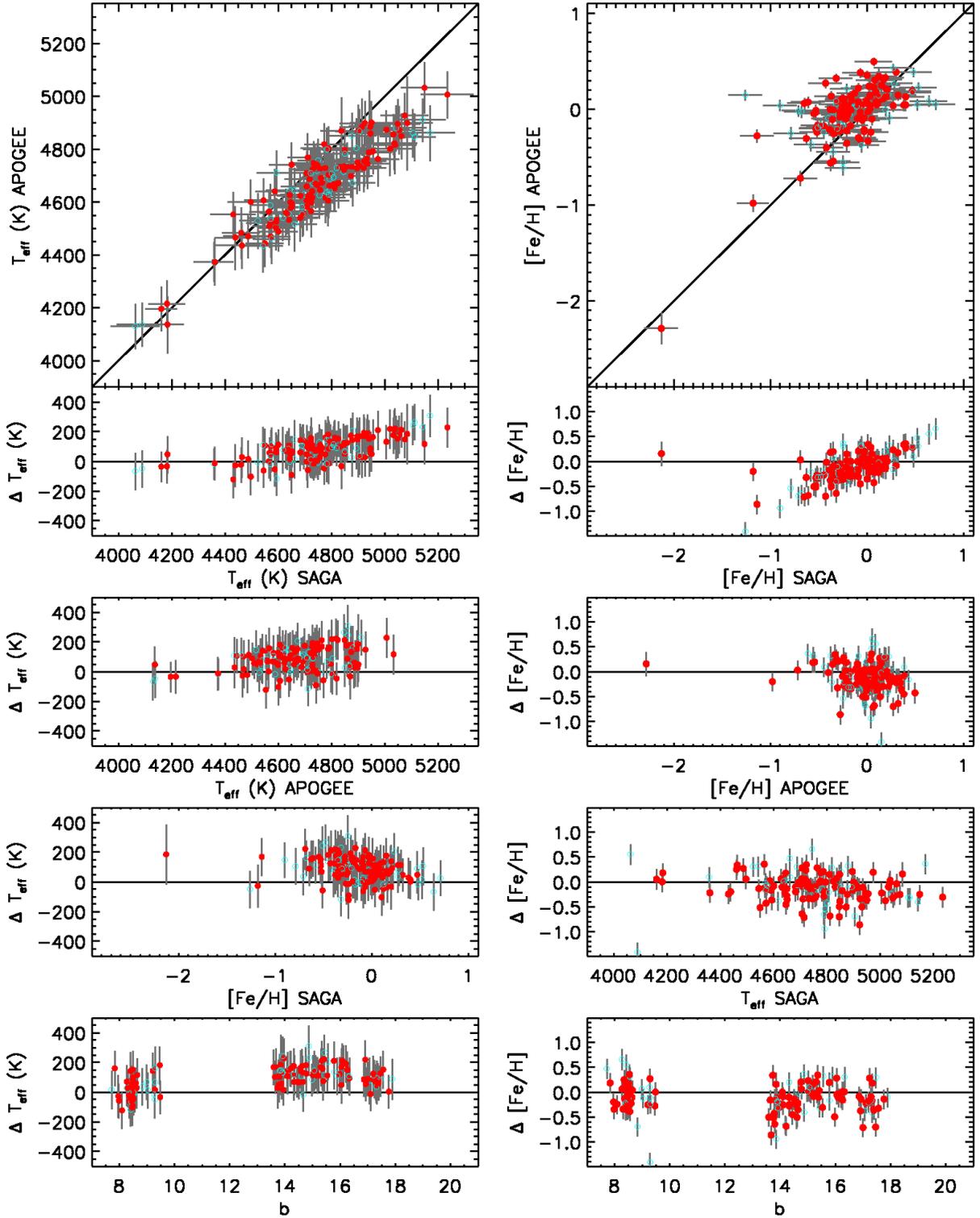}
\caption{Same as Figure \ref{f:kic} but comparing giants in SAGA with APOGEE 
($\Delta=$ SAGA minus APOGEE). 
\label{f:apo}}
\end{figure}

\clearpage

\begin{deluxetable}{ccc}
\tablewidth{0pc}
\tablecaption{SAGA Asteroseismic Catalog Format}
\tablehead{\colhead{Column}   & \colhead{Format} & \colhead{Description}}
\startdata
1 & A8      & KIC ID \\
2 & F6.1    & Frequency of maximum oscillations power $\nu_\mathrm{max}$ ($\mu$Hz) \\
3 & F6.1    & Error in $\nu_\mathrm{max}$ ($\mu$Hz) \\
4 & F7.2    & Large frequency separation $\Delta\nu$ ($\mu$Hz) \\
5 & F7.2    & Error in $\Delta\nu$ ($\mu$Hz) \\
6 & F5.2    & Str\"omgren $\feh$   \\
7 & F5.2    & Error in Str\"omgren $\feh$ \\
8 & F6.0    & $\teff$ from the IRFM \\
9 & F6.0    & Error in $\teff$ \\
10 & F7.2   & Stellar radius ($R_{\odot}$) \\
11 & F7.2   & Upper error in stellar radius \\
12 & F7.2   & Lower error in stellar radius \\
13 & F7.2   & Stellar mass ($M_{\odot}$) \\ 
14 & F7.2   & Upper error in stellar mass \\
15 & F7.2   & Lower error in stellar mass \\
16 & F6.3   & Surface gravity  \\
17 & F6.3   & Upper error in surface gravity \\
18 & F6.3   & Lower error in surface gravity \\
19 & F10.7  & Stellar density ($\rho_{\odot}$) \\
20 & F10.7  & Upper error in stellar density \\
21 & F10.7  & Lower error in stellar density \\
22 & F6.0   & Distance from the Sun (pc) \\
23 & F6.0   & Upper error in distance  \\
24 & F6.0   & Lower error in distance \\
25 & A5     & Asteroseismic classification \\
26 & I1     & Metallicity flag \\
27 & I1     & Str\"omgren photometry flag \\
28 & F5.3   & Str\"omgren $(b-y)$ index \\
29 & F5.3   & Str\"omgren $m_1$ index \\
30 & F5.3   & Str\"omgren $c_1$ index \\
\enddata
\label{t:t1}
\end{deluxetable}

\end{document}